\documentclass[peerreview,onecolumn,draftclsnofoot]{IEEEtran}

\usepackage{amsfonts,amsmath,amssymb,bm,cite,graphicx,hyperref,multirow,subfigure,url}

\begin{document}
\title{Deep Learning-Based Video Coding:\\A Review and A Case Study}
\author{Dong Liu, Yue Li, Jianping Lin, Houqiang Li, Feng Wu
\thanks{The authors are with the CAS Key Laboratory of Technology in Geo-Spatial Information Processing and Application System, University of Science and Technology of China, Hefei 230027, China (e-mail: dongeliu@ustc.edu.cn).}
}

\maketitle
\begin{abstract}
The past decade has witnessed great success of deep learning technology in many disciplines, especially in computer vision and image processing. However, deep learning-based video coding remains in its infancy. This paper reviews the representative works about using deep learning for image/video coding, which has been an actively developing research area since the year of 2015. We divide the related works into two categories: new coding schemes that are built primarily upon deep networks (deep schemes), and deep network-based coding tools (deep tools) that shall be used within traditional coding schemes or together with traditional coding tools. For deep schemes, pixel probability modeling and auto-encoder are the two approaches, that can be viewed as predictive coding scheme and transform coding scheme, respectively. For deep tools, there have been several proposed techniques using deep learning to perform intra-picture prediction, inter-picture prediction, cross-channel prediction, probability distribution prediction, transform, post- or in-loop filtering, down- and up-sampling, as well as encoding optimizations. According to the newest reports, deep schemes have achieved comparable or even higher compression efficiency than the state-of-the-art traditional schemes, such as High Efficiency Video Coding (HEVC) based scheme, for \emph{image} coding; deep tools have demonstrated the compression capability beyond HEVC for \emph{video} coding. However, deep schemes have not yet reached the current height of HEVC for \emph{video} coding, and deep tools remain largely unexplored at many aspects including the tradeoff between compression efficiency and encoding/decoding complexity, the optimization for perceptual naturalness or semantic quality, the speciality and universality, the federated design of multiple deep tools, and so on. In the hope of advocating the research of deep learning-based video coding, we present a case study of our developed prototype video codec, namely Deep Learning Video Coding (DLVC). DLVC features two deep tools that are both based on convolutional neural network (CNN), namely CNN-based in-loop filter (CNN-ILF) and CNN-based block adaptive resolution coding (CNN-BARC). Both tools help improve the compression efficiency by a significant margin. With the two deep tools as well as other non-deep coding tools, DLVC is able to achieve on average 39.6\% and 33.0\% bits saving than HEVC, under random-access and low-delay configurations, respectively. The source code of DLVC has been released for future researches.
\end{abstract}

\begin{IEEEkeywords}
Deep learning, image coding, prediction, transform, video coding.
\end{IEEEkeywords}
\IEEEpeerreviewmaketitle
\section{Introduction}
\subsection{Image/Video Coding}
Image/video coding usually refers to the computing technology that compresses image/video into binary code (i.e. bits) so as to facilitate storage and transmission. The compression may or may not ensure perfect reconstruction of image/video from the bits, which is termed lossless and lossy coding respectively. For natural image/video, the compression efficiency of lossless coding is usually below requirement, so most of efforts are devoted to lossy coding. Lossy image/video coding solutions are evaluated at two aspects: first is the compression efficiency, commonly measured by the number of bits (coding rate), the less the better; second is the incurred loss, commonly measured by the quality of the reconstructed image/video compared with the original image/video, the higher the better.

Image/video coding is a fundamental and enabling technology for computer image processing, computer vision, and visual communication.
The research and development of image/video coding can be dated back to as early as 1960s, much earlier than the appearance of modern imaging, image processing, and visual communication systems.
As an example, Picture Coding Symposium, a prestigious international forum devoted specifically to advancements in image/video coding, started in the year of 1969.
Since then, numerous efforts from both academia and industry have been devoted to this field.

Due to the requirement of interoperability, a series of standards regarding image/video coding have been crafted in the past three decades. In international standardization organizations, ISO/IEC has two experts group namely Joint Photographic Experts Group (JPEG) and Moving Picture Experts Group (MPEG) for standardization of image/video coding technology, while ITU-T has its own Video Coding Experts Group (VCEG). These organizations have published several famous, widely adopted standards, such as JPEG \cite{wallace1992jpeg}, JPEG 2000 \cite{skodras2001jpeg}, H.262 (MPEG-2 Part 2) \cite{tudor1995mpeg}, H.264 (MPEG-4 Part 10 or AVC) \cite{wiegand2003overview}, H.265 (MPEG-H Part 2 or HEVC) \cite{sullivan2012overview}, and so on. At present, H.265/HEVC, which was formally published in 2013, represents the state-of-the-art image/video coding technology.

Along with the progress of video technology, especially the popularization of ultra-high definition (UHD) video, there is an urgent requirement to further increase compression efficiency so as to accommodate UHD video in limited storage and limited transmission bandwidth. Thus, after HEVC, MPEG and VCEG form the Joint Video Experts Team (JVET) to explore advanced video coding technology, and the team developed Joint Exploration Model (JEM) for study. Moreover, since the year of 2018, the JVET team has been working on a new video coding standard, informally called Versatile Video Coding (VVC), as the successor of HEVC. It is anticipated that VVC may improve the compression efficiency by saving around 50\% bits while maintaining the same quality, especially for UHD video, compared to HEVC. Nonetheless, it is worth noting that the improvement of VVC is probably achieved at the cost of multiplicative encoding/decoding complexity.
\subsection{Deep Learning for Image/Video Coding}
The past decade has witnessed the emerging and booming of deep learning, a class of techniques that are increasingly adopted in the hope of approaching the ultimate goal of artificial intelligence \cite{lecun2015deep}. Deep learning belongs to machine learning technology, and has the distinction of its computational models, known as deep artificial neural networks or deep networks for short, which are composed of multiple (usually more than three) processing layers, each layer is further composed of multiple simple but non-linear basic computational units. One benefit of such deep networks is believed to be the capacity for processing data with multiple levels of abstraction, and converting data into different kinds of representations. Note that these representations are not manually designed; instead, the deep network including the processing layers is learned from massive data using a general machine learning procedure. Deep learning eliminates the necessity of handcrafted representations, and thus is regarded useful especially for processing natively unstructured data, such as acoustic and visual signal, whilst processing such data has been a longstanding difficulty in the artificial intelligence field.

Specifically for processing image/video, deep learning using convolutional neural network (CNN) has revolutionized the paradigm in computer vision and image processing. In 2012, Krizhevsky \emph{et al.} \cite{krizhevsky2012imagenet} designed a 8-layer CNN, which won the image classification challenge by a surprisingly low error rate compared with previous works. In 2014, Girshick \emph{et al.} \cite{girshick2014rich} promoted the performance of object detection by a significant margin with the proposed regions with CNN features. Also in 2014, Dong \emph{et al.} \cite{dong2014learning} proposed a 3-layer CNN for single image super-resolution (SR), which outperforms the previous methods in both reconstruction quality and computational speed. In 2017, Zhang \emph{et al.} \cite{zhang2017beyond} presented a deep CNN for image denoising, and demonstrated that a single CNN model may tackle with several different image restoration tasks including denoising, single image SR, and compression artifact reduction, while these tasks had been studied separately for a long while.

Witnessing such successful cases, experts cannot help but ask whether deep learning can benefit image/video coding as well. In history, artificial neural network is not strange to the image/video coding community. From 1980s to 1990s, a number of researches were conducted on neural network-based image coding \cite{dony1995neural,jiang1999image}, but then the networks were shallow and the compression efficiency was not satisfactory. Thanks to the abundance of data, the more and more powerful computing platform, and the development of advanced algorithms, it is now possible to train very deep networks with even more than 1000 layers \cite{he2016deep}. Thus, the exploration of using deep learning for image/video coding is worthy of reconsideration, and indeed has been an actively developing research area since 2015. At present, researches have shown promising results, confirmed the feasibility of deep learning-based image/video coding. Nonetheless, this technology is far from mature and calls for much more research and development efforts.

In this paper, we aim to provide a comprehensive review of the newest reports about deep learning-based image/video coding (until the end of 2018), as well as to present a case study of our developed prototype video codec namely Deep Learning Video Coding (DLVC), so as to make interested readers aware of the status quo. Readers may also refer to \cite{ma2019image} for a recently published review paper about the same theme.

The remainder of this paper is organized as follows. Sections \ref{sec_rev_deepscheme} and \ref{sec_rev_deeptool} provide a review of related works about using deep learning for image/video coding. The related works are divided into two categories, and reviewed in the two sections respectively. The first category is deep schemes, i.e. new coding schemes that are built primarily upon deep networks; the second category is deep tools, i.e. deep network-based coding tools that are embedded into traditional, non-deep coding schemes; a deep tool may either replace its counterpart in the traditional scheme, or be newly added into the scheme. Section \ref{sec_dlvc} presents the case study of our developed DLVC, with all the design details and experimental results. Section \ref{sec_concl} summarizes our perspectives on some open problems for future research, and then concludes this paper. Table \ref{table_abbrev} lists the abbreviations used in this paper.
\begin{table}[tb]
  \centering
  \caption{List of abbreviations}\label{table_abbrev}
\begin{tabular}{l|l}
  \hline
  \textbf{Abbreviation} & \textbf{Remark} \\ \hline
  AVC & Advanced Video Coding, i.e. H.264 \cite{wiegand2003overview} \\
  BARC & block adaptive resolution coding \\
  BD-rate & Bjontegaard's delta-rate \cite{bjontegaard2001calcuation} \\
  BPG & Better Portable Graphics, an image coding format based on HEVC \\
  CNN & convolutional neural network \\
  CTU & coding tree unit \\
  CU & coding unit \\
  DLVC & Deep Learning Video Coding, our developed prototype video codec \\
  GAN & generative adversarial network \\
  HEVC & High Efficiency Video Coding, i.e. H.265 \cite{sullivan2012overview} \\
  HM & HEVC reference software \\
  ILF & in-loop filter \\
  JEM & Joint Exploration Model, a video coding software developed by JVET \\
  JPEG$^1$ & Joint Photographic Experts Group, a group of ISO/IEC \\
  JPEG$^2$ & a standard published by ISO/IEC \cite{wallace1992jpeg} \\
  JVET &  Joint Video Experts Team, a team of MPEG and VCEG \\
  LSTM & long short-term memory \\
  MAE & mean-absolute-error \\
  MC & motion compensation \\
  ME & motion estimation \\
  MPEG & Moving Picture Experts Group, a group of ISO/IEC \\
  MSE & mean-squared-error \\
  MS-SSIM & multi-scale SSIM \\
  PSNR & peak signal-to-noise ratio \\
  QP & quantization parameter \\
  ReLU & rectified linear unit \cite{nair2010rectified} \\
  RNN & recurrent neural network \\
  SR & super-resolution \\
  SSIM & structural similarity \cite{wang2004image} \\
  VCEG & Video Coding Experts Group, a group of ITU-T \\
  VVC & Versatile Video Coding, an incoming video coding standard \\
  \hline
\end{tabular}
\end{table}
\subsection{Preliminaries}
In this paper, we consider coding methods for natural image/video, i.e. the image/video as-is seen by human taken by daily cameras or mobile phones. Although the methods are usually generally applicable, they have been specifically designed for natural image/video, and may not perform that well for other kinds (e.g. biomedical, remote-sensing).

Currently, almost all the natural image/video is in digital format. A grayscale digital image can be denoted by $\bm{x}\in\mathbb{D}^{m\times n}$, where $m$ and $n$ are the number of rows (height) and number of columns (width) of the image, and $\mathbb{D}$ is the definition domain of a single picture element (pixel). For example, $\mathbb{D}=\{0,1,\dots,255\}$ is a common setting, where $|\mathbb{D}|=256=2^8$, thus the pixel value can be represented by an 8-bit integer; accordingly, an uncompressed grayscale digital image has 8 bits-per-pixel (bpp), while compressed bits are definitely less.

A color image is usually decomposed into multiple channels to record the color information. For example, using the RGB color space, a color image can be denoted by $\bm{x}\in\mathbb{D}^{m\times n\times 3}$, where 3 corresponding to three channels--Red, Green, Blue. Since human vision is more sensitive to luminance than to chrominance, the YCbCr (YUV) color space is much more adopted than RGB, and the U and V channels are typically down-sampled to achieve compression. For example, in the so-called YUV420 color format, a color image can be denoted by $\bm{X}=\{\bm{x}_Y\in\mathbb{D}^{m\times n},\bm{x}_U\in\mathbb{D}^{\frac{m}{2}\times \frac{n}{2}},\bm{x}_V\in\mathbb{D}^{\frac{m}{2}\times \frac{n}{2}}\}$.

A color video is composed by multiple color images, called frames, to record the scene at different timestamps. For example, in the YUV420 color format, a color video can be denoted by $\bm{V}=\{\bm{X}_0,\bm{X}_1,\dots,\bm{X}_{T-1}\}$ where $T$ is the number of frames, each $\bm{X}_i=\{\bm{x}^{(i)}_Y\in\mathbb{D}^{m\times n},\bm{x}^{(i)}_U\in\mathbb{D}^{\frac{m}{2}\times \frac{n}{2}},\bm{x}^{(i)}_V\in\mathbb{D}^{\frac{m}{2}\times \frac{n}{2}}\}$. If $m=1080,n=1920,|\mathbb{D}|=2^{10}$, and a video has 50 frames-per-second (fps), then the data rate of the uncompressed video is $1080\times1920\times(10+\frac{10}{4}+\frac{10}{4})\times50=1,555,200,000$ bits-per-second (bps), about 1.555 Gbps. Obviously, the video should be compressed by a ratio of hundreds to thousands before it can be efficiently transmitted over the current wired and wireless networks.

The existing lossless coding methods can achieve a compression ratio of about 1.5 to 3 for natural image, which is clearly below requirement. Thus, lossy coding is introduced to compress more but at the cost of incurring loss. The loss can be measured by the difference between original and reconstructed images, e.g. using mean-squared-error (MSE) for grayscale image:
\begin{equation}
\mbox{MSE}=\frac{||\bm{x}-\bm{x}_{\mbox{rec}}||^2}{m\times n}
\end{equation}
Accordingly, the quality of reconstructed image compared with original image can be measured by peak signal-to-noise ratio (PSNR):
\begin{equation}
\mbox{PSNR}=10\times\log_{10}{\frac{(\max(\mathbb{D}))^2}{\mbox{MSE}}}
\end{equation}
where $\max(\mathbb{D})$ is the maximal value in $\mathbb{D}$, e.g. 255 for 8-bit grayscale image. For color image/video, the PSNR values of Y, U, V are usually separately calculated. For video, the PSNR values of different frames are usually separately calculated and then averaged. There are other quality metrics in replacement of PSNR, such as structural similarity (SSIM) and multi-scale SSIM (MS-SSIM) \cite{wang2004image}.

To compare different lossless coding schemes, it is sufficient to compare the compression ratio, or the resulting rate (bpp, bps, etc.). To compare different lossy coding schemes, it is necessary to take into account both rate and quality. For example, to calculate the relative rates at several different quality levels, and then to average the rates, is a commonly adopted method; the average relative rate is known as Bjontegaard's delta-rate (BD-rate) \cite{bjontegaard2001calcuation}. There are other important aspects to evaluate image/video coding schemes, including encoding/decoding complexity, scalability, robustness, and so on.
\section{Review of Deep Schemes}\label{sec_rev_deepscheme}
In this section we review some representative coding schemes that are built primarily upon deep networks. Generally speaking, there are two approaches for deep image coding schemes, i.e. pixel probability modeling and auto-encoder. These two approaches are combined together in several deep schemes. In addition, we discuss deep video coding schemes and special-purpose coding schemes, where special-purpose schemes are further categorized into perceptual coding and semantic coding.
\subsection{Pixel Probability Modeling}
According to Shannon's information theory \cite{shannon1948mathematical}, the optimal method for lossless coding can reach the minimal coding rate $-\log_2 p(x)$ where $p(x)$ is the probability of the symbol $x$. To reach this target, a number of lossless coding methods have been invented, and arithmetic coding is believed to be among the optimal ones \cite{witten1987arithmetic}. In essence, given a probability $p(x)$, arithmetic coding ensures the coding rate to be as near as possible to $-\log_2 p(x)$ up to rounding error. Thus, the remaining problem is to find out the probability, which is however very difficult for natural image/video as it is of very high dimension.

\begin{figure}[tb]
\centering
   \includegraphics[width=0.3\columnwidth]{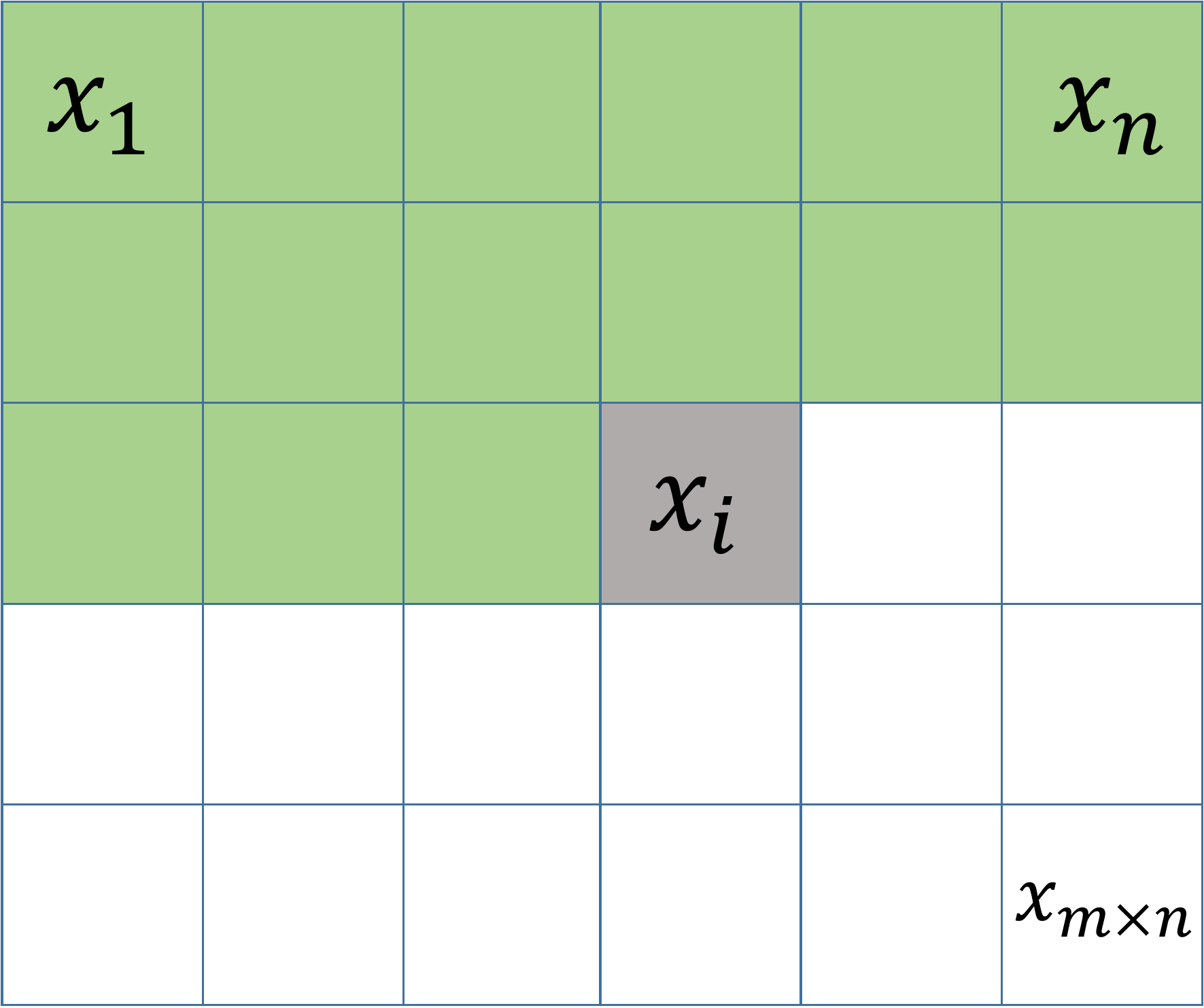}
   \caption{Illustration of a typical predictive coding scheme, where the pixels are encoded/decoded one by one in the raster scan order. For the pixel $x_i$ (marked gray), all the previous pixels (marked green), i.e. above of $x_i$ and left in the same row of $x_i$, can be used as condition to predict the pixel value of $x_i$. The green area is also called the \emph{context} for $x_i$. For simplification, context can be chosen as a subset of the green area.}
\label{fig:conditional_image_generation}
\end{figure}
One way to estimate $p(\bm{x})$, where $\bm{x}$ is an image, is to decompose the image into $m\times n$ pixels and to estimate the probabilities of these pixels one by one (e.g. in the raster scan order). This is a typical \emph{predictive coding} strategy. Note that
\begin{equation}
\label{eqn_conditional_distribution}
p(\bm{x}) = p(x_1)p(x_2|x_1)\cdots p(x_i|x_1,\dots,x_{i-1})\cdots p(x_{m\times n}|x_1,\dots,x_{m\times n-1})
\end{equation}
which is illustrated in Fig. \ref{fig:conditional_image_generation}.
Here the condition for $x_i$ is also called the \emph{context} for $x_i$.
When the image is large, the conditional probability can be difficult to estimate. A simplification is to reduce the range of context, e.g.
\begin{equation}
\label{simple_conditional_distribution}
p(\bm{x}) = p(x_1)p(x_2|x_1)\cdots p(x_i|x_{i-k},\dots,x_{i-1})\cdots p(x_{m\times n}|x_{m\times n-k},\dots,x_{m\times n-1})
\end{equation}
where $k$ is a prechosen constant.

As known, deep learning is good at solving regression and classification problems. Therefore, it has been proposed to estimate the probability $p(x_i|x_1,\dots,x_{i-1})$ given the context $x_1,\dots,x_{i-1}$, using trained deep networks.
This strategy is proposed for other kinds of high-dimensional data in as early as 2000 \cite{bengio2000modeling}, but is applied to image/video until recently.
For example, in \cite{larochelle2011neural}, the probability estimation is considered for binary images, i.e. $x_i\in\{-1,+1\}$, where it suffices to predict a single probability value $p(x_i=+1|x_1,\dots,x_{i-1})$ for each pixel. The paper presents the neural autoregressive distribution estimator (NADE), where a feed-forward network with one hidden layer is used for each pixel, and the parameters are shared across these networks. The parameter sharing also help to speed up the computations for each pixel.
A similar work is presented in \cite{gregor2011learning}, where the feed-forward network also has connections skipping the hidden layer, and the parameters are also shared.
Both \cite{larochelle2011neural} and \cite{gregor2011learning} perform experiments on the binarized MNIST dataset\footnote{The raw MNIST dataset can be accessed at \url{http://yann.lecun.com/exdb/mnist/}.}.
Uria \emph{et al.} \cite{uria2013rnade} extend the NADE to real-valued NADE (RNADE), where the probability $p(x_i|x_1,\dots,x_{i-1})$ is made up by a mixture of Gaussians, and the feed-forward network needs to output a set of parameters for the Gaussian mixture model, instead of a single value in NADE. Their feed-forward network has one hidden layer and parameter sharing, but the hidden layer is equipped with rescaling to avoid saturation, and uses rectified linear unit (ReLU) \cite{nair2010rectified} instead of sigmoid. They also consider mixture of Laplacians rather than Gaussians.
Experiments are conducted on 8$\times$8 natural image patches, where the pixel value is added with noise and converted to real value.
In \cite{uria2014deep}, NADE and RNADE are improved by using different orderings of the pixels as well as using more hidden layers in the network.
In \cite{van2014factoring}, RNADE is improved by enhancing the Gaussian mixture model (GMM) with deep GMM.

Designing advanced networks has been an important theme for improving pixel probability modeling. In \cite{theis2015generative}, multi-dimensional long short-term memory (LSTM) based network is proposed, together with mixtures of conditional Gaussian scale mixtures, a generalization of GMM, for probability modeling. LSTM is a kind of recurrent neural networks (RNNs), and is regarded good at modeling sequential data. The spatial variant of LSTM is used for images. Later in \cite{oord2016pixel}, several different networks are studied, including RNNs and CNNs that are known as PixelRNN and PixelCNN, respectively. For PixelRNN, two variants of LSTM, called row LSTM and diagonal BiLSTM, are proposed, where the latter is specifically designed for images. PixelRNN incorporates residual connections \cite{he2016deep} to help train deep networks with up to 12 layers. For PixelCNN, in order to suit for the shape of context (see Fig. \ref{fig:conditional_image_generation}), \emph{masked} convolutions are proposed. PixelCNN is also as deep as having 15 layers. Compared with previous works, PixelRNN and PixelCNN are more dedicated to natural images: they consider pixels as discrete values (e.g. $0,1,\dots,255$), and predict a multinomial distribution over the discrete values; they deal with color images (in RGB color space); multi-scale PixelRNN is proposed; and they work well on the CIFAR-10 and ImageNet datasets. Quite a number of researches follow the approach of PixelRNN and PixelCNN. In \cite{van2016conditional}, Gated PixelCNN is proposed to improve the PixelCNN, and achieves comparable performance with PixelRNN but with much less complexity. In \cite{salimans2017pixelcnn++}, PixelCNN++ is proposed with the following improvements upon PixelCNN: a discretized logistic mixture likelihood is used rather than a 256-way multinomial distribution; down-sampling is used to capture structures at multiple resolutions; additional short-cut connections are introduced to speed up training; dropout is adopted for regularization; RGB is combined for one pixel. In \cite{chen2017pixelsnail}, PixelSNAIL is proposed, in which casual convolutions are combined with self attention.

Most of the aforementioned works directly model pixel probability. In addition, pixel probability may be modeled as a conditional one upon explicit or latent representations. That says, we may estimate
\begin{equation}
p(\bm{x}|\bm{h})=\prod_{i=1}^{m\times n}p(x_i|x_1,\dots,x_{i-1},\bm{h})
\end{equation}
where $\bm{h}$ is the additional condition. Note also that $p(\bm{x})=p(\bm{h})p(\bm{x}|\bm{h})$, which means the modeling is split into an unconditional and a conditional. For example, in \cite{van2016conditional}, the additional condition can be image class or high-level image representations that are derived by another deep network. In \cite{kolesnikov2016latent}, PixelCNN with latent variables are considered, where the latent variables are derived from the original image: they can be a quantized grayscale version of the original color image, or a multi-resolution image pyramid.

Regarding practical image coding schemes, in \cite{li2018efficient}, a network with \emph{trimmed} convolutions is adopted to predict probabilities for binary data, while a 8-bit grayscale image with the size of $m\times n$ is converted to a binary cube with the size of $m\times n\times8$ to be processed by the network. The network is similar to PixelCNN but is of three dimension. The trimmed convolutional network-based arithmetic encoding (TCAE) is reportedly better than the previous non-deep lossless coding schemes, such as TIFF, GIF, PNG, JPEG-LS, and JPEG 2000-LS; on the Kodak image set\footnote{The Kodak image set can be accessed at \url{http://www.r0k.us/graphics/kodak/}.}, TCAE achieves a compression ratio of 2.00. Differently, in \cite{ahanonu2018lossless}, CNN is used in the wavelet transform domain rather than the pixel domain, i.e. CNN is to predict wavelet detail coefficients from coefficients within neighboring subbands.

For video coding, in \cite{kalchbrenner2016video}, PixelCNN is generalized to video pixel network (VPN) for the pixel probability modeling of video. VPN is composed of CNN encoders (for previous frames to predict the current frame) and PixelCNN decoders (for prediction inside the current frame). CNN encoders preserve at all layers the spatial resolution of input frames to maximize representational capacity. Dilated convolutions are adopted to enlarge receptive fields and better capture global motion. The outputs of the CNN encoders are combined over time with a convolutional LSTM that also preserves the resolution. The PixelCNN decoders use masked convolutions and adopt multinomial distributions over discrete pixel values. VPN is experimented on the Moving MNIST and Robotic Pushing datasets.

In addition, Schiopu \emph{et al.} \cite{schiopu2018cnn} investigate a lossless image coding scheme, where they use CNN to predict pixel value rather than its distribution. The predicted value is subtracted from the true pixel value, resulting in residue that is then coded. In addition, they consider the adaptive selection among the CNN predictor and some non-CNN predictors.
\subsection{Auto-Encoder}\label{transform_coding_scheme}
Auto-encoder originates from the well-known work of Hinton and Salakhutdinov \cite{hinton2006reducing}, which trains a network for dimensionality reduction and the network consists of encoding part and decoding part. The encoding part converts an input high-dimension signal to its low-dimension representation, and the decoding part recovers (not perfectly) the high-dimension signal from the low-dimension representation. Auto-encoder enables automated learning of representations and eliminates the need of hand-crafted features, which is also believed to be one of the most important advantages of deep learning.

It seems quite straightforward to adopt the auto-encoder network for lossy image coding: the encoder and decoder are trained out, and we just need to encode the learned representation. However, the traditional auto-encoder is not optimized for compression, and directly using a trained auto-encoder is not efficient \cite{watkins2017image}.
When we consider the compression requirement, there are several challenges: First, the low-dimension representation shall be quantized then coded, but the quantization step is not differentiable, making a difficulty to train the network. Second, lossy coding is to achieve a better tradeoff between rate and quality, so the rate shall be taken into account when training the network, but the rate is not easy to calculate or estimate. Third, a practical image coding scheme needs to consider variable rate, scalability, encoding/decoding speed, interoperability, and so on. In response to these challenges, a number of researches have been conducted especially in recent years.

\begin{figure}[tb]
\centering
   \includegraphics[width=0.6\columnwidth]{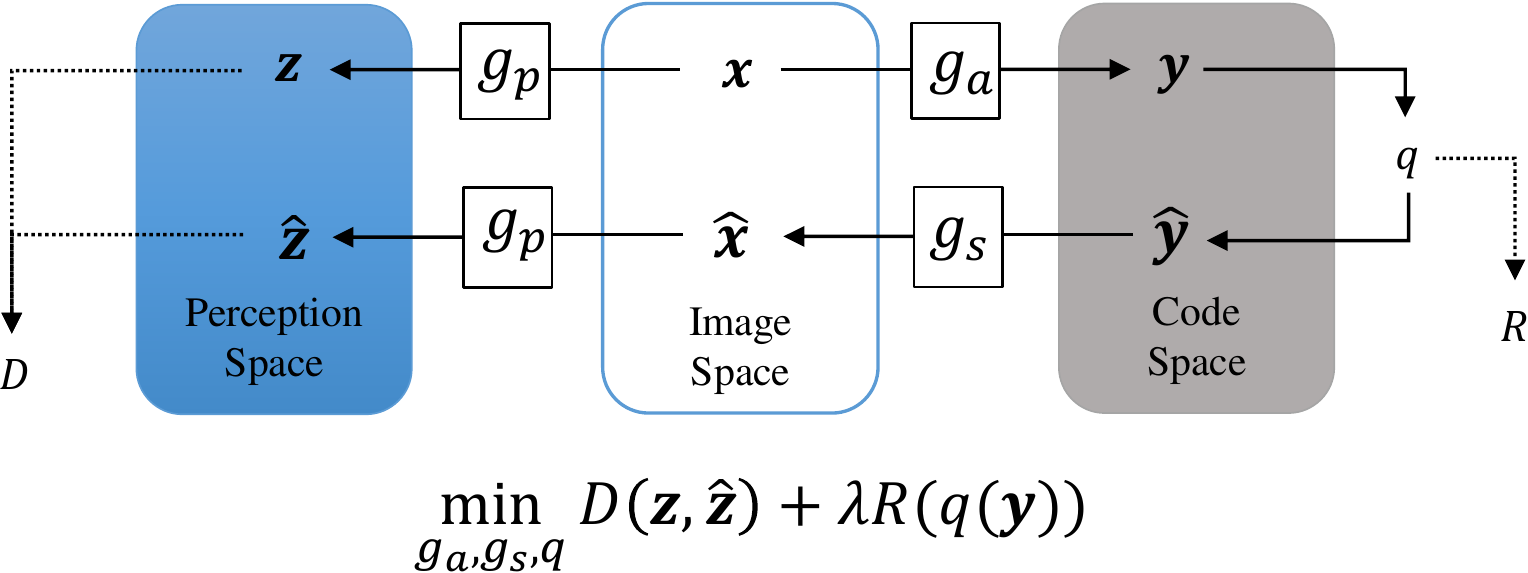}
   \caption{Illustration of a typical transform coding scheme. The original image $\bm{x}$ is transformed by an \emph{analysis} function $g_a$ to achieve the code $\bm{y}$. The code $\bm{y}$ is quantized (denoted by $q$) and compressed into bits. The number of bits is used to measure the coding rate ($R$). The quantized code $\hat{\bm{y}}$ is then inversely transformed by a \emph{synthesis} function $g_s$ to achieve the reconstructed image $\hat{\bm{x}}$. Both of $\bm{x}$ and $\hat{\bm{x}}$ are further transformed by a same \emph{perceptual} function $g_p$, resulting in $\bm{z}$ and $\hat{\bm{z}}$, respectively. The difference between $\bm{z}$ and $\hat{\bm{z}}$ is used to measure the distortion ($D$).}
\label{fig:nonlinear_coding_framework}
\end{figure}
A conceptual illustration of auto-encoder-based image coding scheme is shown in Fig. \ref{fig:nonlinear_coding_framework}, which is a typical \emph{transform coding} strategy. The original image $\bm{x}$ is transformed to $\bm{y}=g_a(\bm{x})$, and $\bm{y}$ is quantized then coded. The decoded $\hat{\bm{y}}$ is inversely transformed to $\hat{\bm{x}}=g_s(\hat{\bm{y}})$. Considering the tradeoff between rate and quality, we can train the network to minimize the joint rate-distortion cost $D+\lambda R$ where $D$ is calculated or estimated as the difference between $\bm{x}$ and $\hat{\bm{x}}$ (note that the difference may be calculated or estimated in a perception space), $R$ is calculated or estimated from the quantized code, and $\lambda$ is the Lagrange multiplier. All of the existing researches follow this scheme more or less and differ in their network structure and loss function.

For the network structure, RNNs and CNNs are the widely used two categories. The most representative works include:
\begin{itemize}
  \item Toderici \emph{et al.} \cite{toderici2015variable} propose a general framework for variable rate image compression. They use binary quantization to generate codes, and do not consider the rate during training, i.e. the loss is only end-to-end distortion, measured by MSE. Their framework indeed provides a scalable coding functionality, where RNN (specifically LSTM) with convolutional and deconvolutional layers is reported to perform well. They provide results on a large-scale dataset of 32$\times$32 thumbnails. Later, Toderici \emph{et al.} \cite{toderici2017full} propose an improved version, where they use a neural network like PixelRNN \cite{oord2016pixel} to compress the binary codes; they also introduce a new gated recurrent unit (GRU) inspired by the residual network (ResNet) \cite{he2016deep}. They report better results than JPEG on the Kodak image set using MS-SSIM as quality metric. Johnston \emph{et al.} \cite{johnston2017improved} further improve the RNN-based method by introducing hidden-state priming into RNN, using an SSIM-weighted loss function, and enabling spatially adaptive bitrates. They achieve better results than BPG on the Kodak image set using MS-SSIM. Covell \emph{et al.} \cite{covell2017target} enable spatially adaptive bitrates by training stop-code tolerant RNNs.
  \item Ball{\'e} \emph{et al.} \cite{balle2016end} propose a general framework for rate-distortion optimized image compression. They use multiary quantization to generate integer codes and consider the rate during training, i.e. the loss is the joint rate-distortion cost, where distortion can be MSE or others. To estimate the rate, they use adding a random noise to replace the quantization during training, and use the differential entropy of the noisy ``code'' as a proxy for the rate. As for the network structure, they use the generalized divisive normalization (GDN) transform, which consists of a linear mapping (matrix multiplication) followed by a nonlinear parametric normalization; the effectiveness of the proposed GDN for image coding is verified in \cite{balle2018efficient}. Later, Ball{\'e} \emph{et al.} \cite{Balle2016} propose an improved version, where they use 3 convolutional layers each followed by down-sampling and a GDN operation to implement the transform; accordingly, the use 3 layers of inverse GDN + up-sampling + convolution to implement the inverse transform. In addition, they design an arithmetic coding method to compress the integer codes. They report better results than JPEG and JPEG 2000 on the Kodak image set using MSE as quality metric. Furthermore, Ball{\'e} \emph{et al.} \cite{balle2018variational} improve their scheme by incorporating a scale hyper-prior into the auto-encoder, which is inspired by the variational auto-encoder \cite{kingma2013auto}. They use another transform $h_a$ to convert $\bm{y}$ into $\bm{w}=h_a(\bm{y})$, quantize and encode $\bm{w}$ (transmitted as side information), and use another inverse transform $h_s$ to convert the decoded $\hat{\bm{w}}$ into the estimated standard deviation of the quantized $\hat{\bm{y}}$, which is then used during the arithmetic coding of $\hat{\bm{y}}$. On the Kodak image set and using PSNR as quality metric, their method is only slightly worse than BPG.
\end{itemize}

Besides \cite{balle2016end}, several works also concentrate on dealing with the non-differentiable quantization and/or the estimation of rate.
Theis \emph{et al.} \cite{theis2017lossy} adopt a very simple work-around for quantization: quantization is performed as usual in the forward pass, but the gradients are directly passed through the quantization layer in the backward pass. Surprisingly this work-around works well.
In addition, they replace the rate with an upper bound that is differentiable.
Dumas \emph{et al.} \cite{dumas2017image} consider a stochastic winner-take-all mechanism, where the entries in $\bm{y}$ with the largest absolute values are kept and the other entries are set to 0; then the entries are uniformly quantized and compressed.
Agustsson \emph{et al.} \cite{agustsson2017soft} propose a soft-to-hard vector quantization scheme, where they use a soft quantization (i.e. assigning a representation to multiple codes with different membership values) rather than hard quantization (i.e. assigning a representation to only one code) during training, and they adopt an annealing process to let the soft quantization approach the hard quantization gradually. Note that their scheme takes advantage of vector quantization while other works usually adopt scalar quantization.
Li \emph{et al.} \cite{li2017learning} introduce an importance map for rate estimation, where the importance map is quantized to a mask and the mask decides how many bits are kept at each location, thus the sum of the importance map can be used as a rough estimate of the coding rate.

Besides \cite{toderici2015variable}, several works also consider the functionality of variable rate with less or no training for different rates. In \cite{theis2017lossy}, scale parameters are introduced and a pretrained auto-encoder is fine-tuned for different rates. In \cite{dumas2018autoencoder}, a unique learned transform is proposed, together with variable quantization step for different rates.
In \cite{cai2018efficient}, a multi-scale decomposition transform is trained and optimized for all scales; and rate allocation algorithms are provided to determine the optimal scale of each image block for either a target rate or a target quality factor.
Besides, scalable coding is considered in \cite{zhang2018learned} differently from that in \cite{toderici2015variable}. In \cite{zhang2018learned}, an image is decomposed into multiple bit-planes, which are transformed and quantized in parallel; bidirectional assembling gated units are proposed to reduce the correlation between different bit-planes.

Several works consider advanced network structures and different loss functions. Theis \emph{et al.} \cite{theis2017lossy} adopt a sub-pixel structure for computational efficiency. Rippel and Bourdev \cite{rippel2017real} present a pyramid decomposition followed by inter-scale alignment network, which is lightweight and runs in real-time. They also use a discriminator loss in addition to the reconstruction loss.
Snell \emph{et al.} \cite{snell2017learning} use the MS-SSIM as loss function instead of MSE or mean-absolute-error (MAE) to train auto-encoders, and they find that MS-SSIM is better calibrated to perceptual quality.
Zhou \emph{et al.} \cite{zhou2018variational} use deeper networks for encoder/decoder and a separate network for post-processing at the decoder side. They also replace the Gaussian model in \cite{balle2018variational} with the Laplacian model.

As mentioned before, pixel probability modeling represents predictive coding and auto-encoder represents transform coding. These two strategies can be combined for higher compression efficiency.
Mentzer \emph{et al.} \cite{mentzer2018practical} propose a practical lossless image coding scheme, where they use auto-encoders at multiple levels to learn the condition for pixel probability modeling.
Mentzer \emph{et al.} \cite{mentzer2018conditional} integrate pixel probability modeling (a 3D PixelCNN) into auto-encoder so as to estimate the coding rate and to train the PixelCNN and the auto-encoder jointly.
Baig \emph{et al.} \cite{baig2017learning} introduce partial-context image inpainting into the variable rate compression framework \cite{toderici2015variable}, which is actually to predict a block from the block's context, assuming the blocks are encoded/decoded one by one in the raster scan order (similar to what is shown in Fig. \ref{fig:conditional_image_generation} but at the block level). The prediction signal is added onto the network output signal to achieve $\hat{\bm{x}}$, i.e. the transform coding network deals with the prediction residues.
Minnen \emph{et al.} \cite{minnen2018spatially} additionally consider rate allocation among the blocks.
Similarly but in a different manner, Minnen \emph{et al.} \cite{minnen2018joint} improve upon \cite{balle2018variational} by augmenting the hyper-prior with the context, i.e. they use not only $\hat{\bm{w}}$ but also the context to predict the probability of each entry of $\hat{\bm{y}}$. Their method outperforms BPG on the Kodak image set and using PSNR as quality metric, which represents the state of the art by the end of 2018.
Lee \emph{et al.} \cite{lee2018context} introduce the context adaptive entropy model into the hyper-prior $\hat{\bm{w}}$.

Moreover, Cheng \emph{et al.} \cite{cheng2018deep} apply principle component analysis on the learned representation, which is virtually a second transform.
\subsection{Video Coding}
Starting from 2017, a few researches have been reported for deep video coding schemes. Compared to image coding, video coding calls for efficient methods to remove the inter-picture redundancy. Inter-picture prediction is then an important issue in these researches. Motion estimation and compensation is widely adopted, but is implemented by trained deep networks until recently.

Chen \emph{et al.} \cite{chen2017deepcoder} seems the first to report a video coding scheme by using trained deep networks as auto-encoders. Specifically, they divide video frames into 32$\times$32 blocks and for each block they choose one from two modes: intra coding or inter coding. If using intra coding, there is an auto-encoder to compress the block. If using inter coding, then they perform motion estimation and compensation using the traditional method, and input the residues to another auto-encoder. For both auto-encoders, the encoded representations are directly quantized and coded by the Huffman method. This scheme is quite rough and does not compete H.264.

Wu \emph{et al.} \cite{wu2018video} propose a video coding scheme with image interpolation, where the key frames (I frames) are first compressed by the deep image coding scheme in \cite{toderici2017full}, and the remaining frames (B frames) are then compressed in a hierarchical order. For each B frame, two compressed frames (either I frames or previously compressed B frames) before and after are used to ``interpolate'' the current frame: the motion information is used to warp the two compressed frames (i.e. motion compensation), and then the two warped frames are sent as side information to a variable rate image coding scheme that processes the current frame. The scheme is reported to perform on par with H.264.

Chen \emph{et al.} \cite{chen2019learning4} propose another video coding scheme with the so-called PixelMotionCNN. In their scheme, frames are compressed in the temporal order, and each frame is divided into blocks that are compressed in the raster scan order. Before one frame is compressed, the previous two compressed frames are used to ``extrapolate'' the current frame. When a block is to be compressed, the extrapolated frame together with the block's context are sent to the PixelMotionCNN to generate a prediction signal for the current block, then the prediction residues are compressed by the variable rate image coding scheme in \cite{toderici2017full}. This scheme also performs on par with H.264.

Lu \emph{et al.} \cite{lu2019dvc} propose a real end-to-end deep video coding scheme, which can be viewed as a ``deepened'' version of the traditional video coding schemes. Specifically in their scheme, for each frame to be compressed, an optical flow estimation module is used to obtain the motion information between the frame and the previous compressed frames. Motion compensation is also performed by a trained network, to generate a prediction signal for the current frame. For the prediction residues and the motion information, two auto-encoders are used to compress them respectively. The entire network is jointly optimized with a single loss function, i.e. the joint rate-distortion cost. This scheme reportedly achieves better compression efficiency than H.264, and even outperforms HEVC (x265 encoder) when evaluated with MS-SSIM.

Rippel \emph{et al.} \cite{rippel2018learned} present the to-date most sophisticated deep video coding scheme, which inherits and extends a deepened version of the traditional video coding schemes. Their scheme has the following new features: (1) only one auto-encoder to compress motion information and prediction residues simultaneously; (2) a state that is learned from the previous frames and updated recursively; (3) motion compensation with multiple frames and multiple optical flows; (4) a rate control algorithm. This scheme is reported to outperform HEVC reference software (HM) when evaluated with MS-SSIM.

By the end of 2018, we do not observe any report that a deep video coding scheme can outperform HM when evaluated with PSNR, which seems a hard mission.
\subsection{Special-Purpose Coding}
Most of the researches about deep schemes concern image/video coding for \emph{signal fidelity}, i.e. to minimize the distortion between original and reconstructed image/video subject to a given rate, where the distortion can be defined as MSE or other differences. However, if we do not concern the fidelity, we may instead care about the perceptual naturalness of the reconstructed image/video, or the utility of the reconstructed image/video in semantic analysis tasks. The latter two kinds of quality metrics are termed \emph{perceptual naturalness} and \emph{semantic quality}. There have been a few works that tailor image/video coding for these quality metrics.
\subsubsection{Perceptual Coding}
\label{Perceptual}
Since the boom of generative adversarial network (GAN) \cite{goodfellow2014generative}, deep networks are known to be capable in generating perceptually natural images. Leveraging this capability at the decoder side can surely improve the perceptual quality of decoded images. Different from the generator in normal GANs, the decoder should also ensure the decoded images to be similar to original images, which raises a problem of controlled generation and the encoder actually provides the control signal in the coded bits.

Inspired by the variational auto-encoder (VAE) \cite{kingma2013auto}, Gregor \emph{et al.} \cite{gregor2015draw} propose Deep Recurrent Attentive Writer (DRAW) for image generation, which extends the traditional VAE by using RNNs as encoder and decoder. Unfolding the encoder RNN produces a series of latent representations. Then, Gregor \emph{et al.} \cite{gregor2016towards} introduce convolutional DRAW, and observe that it is able to transform an image into a series of increasingly detailed representations, ranging from global conceptual aspects to low level details. Thus, they suggest a conceptual compression scheme, whose one benefit is to achieve plausible reconstruction images at very low bit rates.

It has been realized that perceptual naturalness can be evaluated by the discriminator in GAN \cite{blau2018perception}. Several researches are devoted to deep coding schemes for perceptual quality using the discriminator loss solely or jointly with MSE or other losses.
For example, Santurkar \emph{et al.} \cite{santurkar2018generative} propose the so-called generative compression schemes for both image and video. For image, they first train a canonical GAN, then they use the generator as the decoder, fix it, and train the encoder to minimize a sum of MSE and feature loss. For video, they reuse the encoder and decoder trained for image, transmit only a few frames, and restore the other frames at the decoder side via interpolation. Their schemes are able to achieve very high compression ratio.
Kim \emph{et al.} \cite{kim2018adversarial} build a new video compression scheme, where a few key frames are normally compressed (by H.264) and the other frames are extremely compressed. Indeed, edges are extracted from the down-sampled non-key frames and transmitted. At the decoder side, the key frames are firstly reconstructed, then edges are similarly extracted from them. A conditional GAN is trained with the reconstructed key frames where edge is the condition. Then the conditional GAN is used to generate the non-key frames. Again, their scheme performs well at very low bit rates.
\subsubsection{Semantic Coding}
\label{Semantic}
A few researches have been conducted on deep coding schemes that preserve the semantic information or concern the semantic quality.

Agustsson \emph{et al.} \cite{agustsson2018generative} present a GAN-based image compression scheme for extremely low bit rates. The scheme combines auto-encoder and GAN, collapsing the decoder and the generator into one. In addition, a semantic label map can be used as an additional input to the encoder, and as a condition for the discriminator. It is reported that the proposed scheme reconstructs images with higher semantic quality, in the sense that the semantic segmentation on these images is more accurate than that on BPG-compressed images at the same rate.

Luo \emph{et al.} \cite{luo2018deepsic} propose a concept of deep semantic image compression (DeepSIC), which incorporates the semantic information (e.g. classes) into the coded bits. There are two versions of DeepSIC, both based on auto-encoder. In the one version, the semantic information is extracted from the representation $\bm{y}$, and encoded into the bits. In the other version, the semantic information is not encoded, but extracted at the decoder side from the quantized representation $\hat{\bm{y}}$.
Torfason \emph{et al.} \cite{torfason2018towards} investigate performing semantic analysis tasks (classification and semantic segmentation) from the quantized representation rather than from the reconstructed image. That says, the decoding process is omitted. They show that the classification and segmentation accuracy values are very close between the representation and the image, but the computational complexity is reduced significantly.
Zhang \emph{et al.} \cite{zhang2017deep} study a deep image coding scheme for simultaneous compression and retrieval. Their motivation is that the coded bits can be used not only for reconstructing image but also for retrieving similar images \emph{without decoding}. They use an auto-encoder to compress image into bits, and use a revised classification network to extract binary features. Then they combine the two parts of bits, and fine-tune the feature extraction network for image retrieval. Their results indicate that at the same rate, the reconstructed images are better than JPEG-compressed ones, and the retrieval accuracy improves due to the fine-tuning.

Akbari \emph{et al.} \cite{akbari2018dsslic} design a scalable image coding scheme where the coded bits consist of three layers. The first layer is the semantic segmentation map coded losslessly. The second layer is a down-sampled version of the original image also coded losslessly. With the first two layers, a network is trained to predict the original image and the prediction residues are coded by BPG as the third layer. This scheme is reported to outperform BPG on the Kodak image set when evaluated with PSNR and MS-SSIM.

Chen and He \cite{chen2019learning} consider deep coding for facial images with semantic quality metric instead of PSNR or perceptual quality. For this purpose, their loss function has three parts: MAE, discriminator loss, and a semantic loss, where the semantic loss is to project the original and reconstructed images into a compact Euclidean space through a learned transformation, and to calculate the Euclidean distance between them. Accordingly, their scheme performs very well when evaluated with face verification accuracy at the same rate.
\section{Review of Deep Tools}\label{sec_rev_deeptool}
\begin{figure}[tb]
\centering
   \includegraphics[width=1.0\columnwidth]{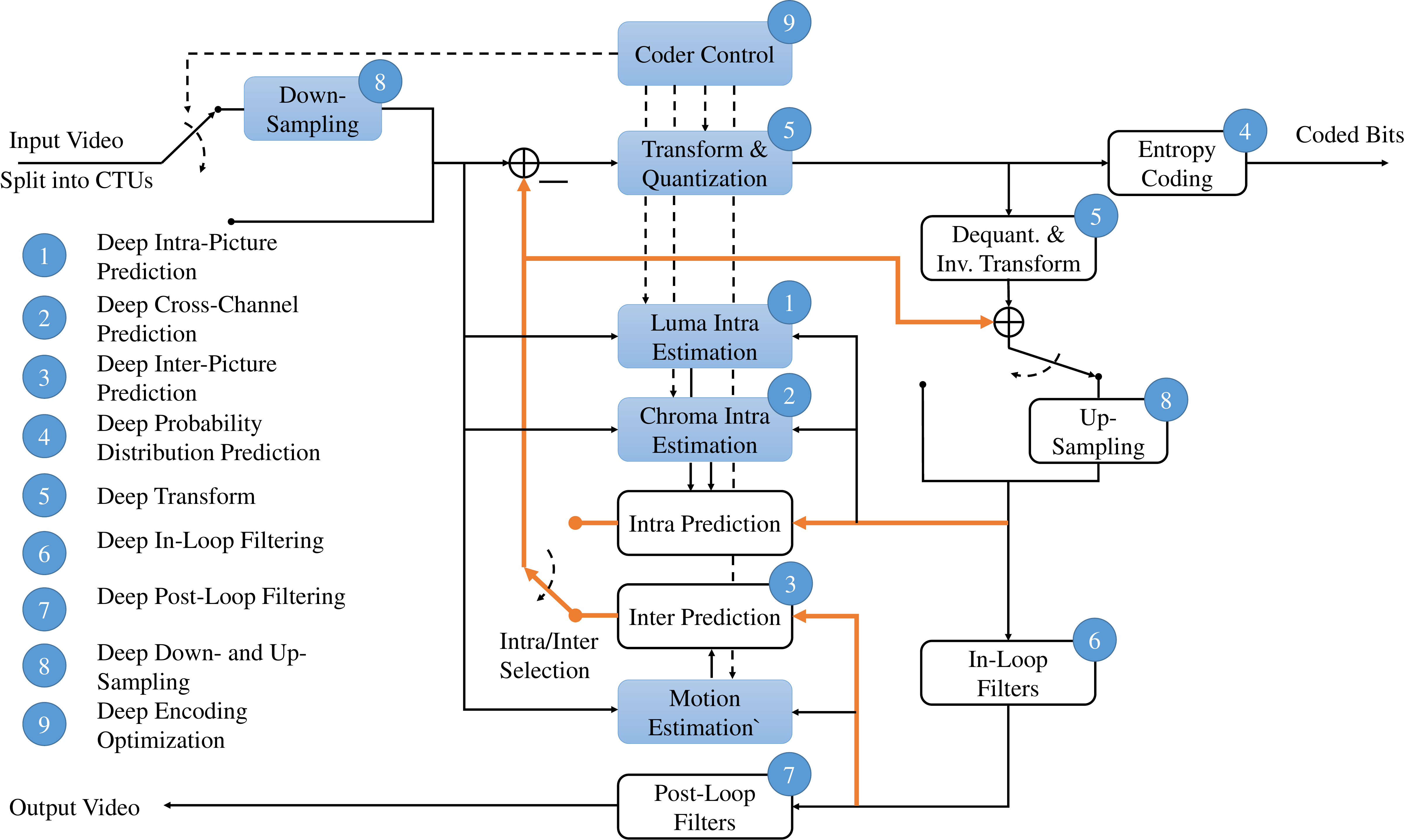}
   \caption{Illustration of a traditional hybrid video coding scheme as well as the locations of deep tools inside the scheme. Note that the yellow lines indicate the flow of prediction, and the blue boxes indicate the tools that are used at the encoder side only.}
\label{fig:deep_tools}
\end{figure}
In this section we review some representative works about using trained deep networks as tools within the traditional coding schemes or together with traditional coding tools. Generally speaking, the traditional video coding schemes adopt a hybrid coding strategy, i.e. a combination of predictive coding and transform coding. As depicted in Fig. \ref{fig:deep_tools}, an input video sequence is divided into pictures, pictures are divided into blocks (the largest block is called CTU, which can be divided into smaller CUs, in HEVC \cite{sullivan2012overview}), and blocks are divided into channels (i.e. Y, U, V). The pictures/blocks/channels are compressed in a predefined order, and the previously compressed ones can be used to predict the following ones, which is known as intra-picture prediction (between blocks), cross-channel prediction (between channels), and inter-picture prediction (between pictures), respectively. The prediction residues are then transformed and quantized and entropy coded to achieve the final bits. Some auxiliary information such as block partition and prediction mode is also entropy coded into the bits (not shown in the figure).
Probability distribution prediction is used in the entropy coding step.
Since the quantization step loses information and may cause artifacts, filtering is proposed to enhance the reconstructed video, which may be performed in-loop (before predicting the next picture) or out-of-loop (before output).
In addition, to reduce the data volume, the pictures/blocks/channels may be down-sampled before being compressed, and up-sampled afterwards.
Finally, the encoder needs to control the different modules and combine them to achieve a tradeoff between coding rate, quality, and computational speed. Encoding optimization is an important theme in practical coding systems.

Trained deep networks can act as almost all of the modules shown in Fig. \ref{fig:deep_tools}, where we have indicated different locations for deep tools. In the following, we will review the works about deep tools according to where they are used in the entire scheme.
\subsection{Intra-Picture Prediction}
Intra-picture prediction, or intra prediction for short, is a tool to predict between blocks inside the same picture. H.264 introduces intra prediction with several predefined prediction modes, such as DC prediction and extrapolation along different directions \cite{wiegand2003overview}. The encoder can choose a prediction mode for each block and signal the choice to the decoder. To decide mode, it is a common strategy to compare the coding rate and distortion of different modes and to select the mode with the minimal rate-distortion cost. In HEVC, more prediction modes are introduced \cite{sullivan2012overview}.

\begin{figure}[tb]
\centering
   \includegraphics[width=0.8\columnwidth]{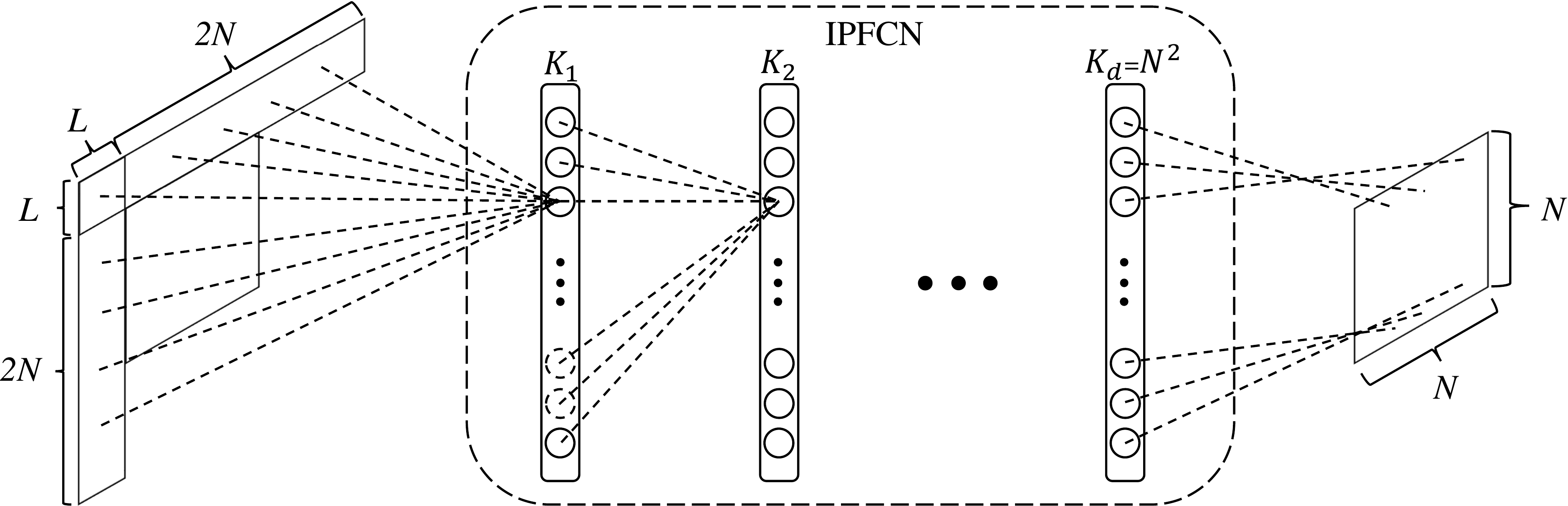}
   \caption{Illustration of a fully connected network for intra prediction (IPFCN).}
\label{fig:IPFCN}
\end{figure}
Li \emph{et al.} \cite{li2018fully} propose a fully connected network for intra prediction that is depicted in Fig. \ref{fig:IPFCN}. For the current $N\times N$ block, they use $L$ rows above and $L$ columns to the left, in total $4NL + L^2$ pixels as context. They use an image set known as the New York City Library to generate training data, in which the raw image is compressed at different quantization parameters (QPs). When training the network, they investigate two strategies: the first is to train a single model with all training data, and the second is to split the training data into two groups by considering the HEVC prediction modes, and to train two models respectively. The strategy of two models turns out better for compression. They integrate the trained networks as new prediction modes along with the HEVC modes. They report around 3\% BD-rate reduction than HM.

Pfaff \emph{et al.} \cite{pfaff2018neural} also adopt fully connected network for intra prediction, but propose to train multiple networks as different prediction modes. In addition, they propose to train a separate network whose input is also the block's context but output is the predicted likelihood of different modes. Moreover, they propose to use a different transform for each of the network-based prediction modes. Their reported performance is high: around 6\% BD-rate reduction than an improved version of HM (with advanced block partitioning).

Hu \emph{et al.} \cite{hu2018progressive} devise a progressive spatial RNN for intra prediction. Different from the above works, they propose to leverage the sequential modeling capacity of RNN to generate prediction progressively from the context to the block. In addition, they suggest the use of sum-of-absolute-transformed-difference (SATD) as the loss function and argue that SATD correlates better to the rate-distortion cost.

Cui \emph{et al.} \cite{cui2018convolutional} consider a CNN for intra prediction, or more specifically, intra prediction refinement. They use the HEVC prediction modes to generate prediction, and then use a trained CNN to refine the prediction. Note that the CNN has not only the HEVC prediction but also the context as its input. This method seems achieving only marginal gain.
\subsection{Inter-Picture Prediction}
Inter-picture prediction, or inter prediction for short, is a tool to predict between video frames so as to remove the redundancy along the temporal dimension. Inter prediction is the kernel of video coding and it largely decides the compression efficiency of a video coding scheme. In the traditional video coding schemes, inter prediction is mostly fulfilled by block-level motion estimation (ME) and motion compensation (MC). Given a reference frame and a block to be coded, ME is to find the location in the reference frame where the content is the most similar to that inside the to-be-coded block, and MC is to retrieve the content at the found location so as to predict the block. Many techniques have been proposed to improve block-level ME and MC, such as using multiple reference frames, bi-directional inter prediction (i.e. using two reference frames jointly), fractional-pixel ME and MC, and so on.

Inspired by the multiple reference frames, Lin \emph{et al.} \cite{lin2018generative} propose a new inter prediction mechanism by extrapolating the multiple reference frames. Specifically they adopt a Laplacian pyramid of GANs to extrapolate a frame from the previously compressed four frames. This extrapolated frame serves as another reference frame. They report around 2\% BD-rate reduction than HM.

Inspired by the bi-directional inter prediction, Zhao \emph{et al.} \cite{zhao2018enhanced} propose a method to enhance the prediction quality. The previous bi-directional prediction simply computes a linear combination of two prediction blocks. They propose to employ trained CNN to combine the two prediction blocks in a nonlinear and data-driven manner.

Inspired by the fractional-pixel ME and MC, a number of researches are conducted on the fractional-pixel interpolation problem, which aims at generating imaginary pixels at fractional locations on the reference frame because the motion between two frames is not aligned to integer pixels.
Here, a major difficulty is how to prepare training data because fractional pixels are imaginary.
Yan \emph{et al.} \cite{yan2017convolutional} propose to use a CNN for half-pixel interpolation, where they suggest a method that blurs a high-resolution image and then samples pixels from the blurred image: odd locations as integer pixels and even locations as half pixels.
This method is inherited in \cite{liu2019one}, where the authors analyze the effect of different blurring degrees.
Zhang \emph{et al.} \cite{zhang2017learning} propose another method, which formulates the fractional interpolation as a resolution enhancement problem. Thus, they down-sample high-resolution images to achieve training data.
Yan \emph{et al.} \cite{yan2018convolutional} consider a different formulation, treating the fractional-pixel MC as an inter-picture regression problem. They use video sequences to retrieve training data, where they rely on the fractional-pixel ME to align different frames, and use reference frame as integer pixels and current frame as fractional pixels.
Yan \emph{et al.} \cite{Yan2018} further discover a key characteristic of the fractional interpolation problem, namely its invertibility: if fractional pixels can be interpolated from integer pixels, then integer pixels should also be interpolated from fractional pixels. Based on the invertibility, they propose an unsupervised manner to train CNN for half-pixel interpolation.

In addition to the improvements of inter prediction methods, another approach is considered where intra and inter predictions are combined. Specifically, the generation of prediction signal is based on not only reference frame but also context in current frame. For example, Huo \emph{et al.} \cite{Huo2018} propose to use a trained CNN to refine the inter prediction signal. They find that using the context of the to-be-predicted block can improve the prediction quality. Similarly, Wang \emph{et al.} \cite{wang2018neural} also refine the inter prediction signal by a CNN, where the CNN inputs include the inter prediction signal, the context of the current block, and the context of the inter prediction block.
\subsection{Cross-Channel Prediction}
Cross-channel prediction is to predict between different channels. In the YUV format, the luma channel (Y) is usually coded before the chroma channels (U and V). Thus, it is possible to predict U from Y, and to predict V from Y and U. A traditional method, known as Linear Model (LM), is intended for cross-channel prediction. The key idea of LM is that chroma can be predicted from luma using a linear function, but the coefficients of the function is not transmitted; instead, they are estimated from the context by performing a linear regression. The linear assumption seems over simplified.

Baig and Torresani \cite{baig2017multiple} investigate colorization for image compression. Colorization is to predict chroma from luma, which is an ill-posed problem because one luma value can correspond to multiple chroma values. Accordingly, they propose a tree-structured CNN, which is able to generate multiple predictions (called multiple hypotheses) given one grayscale image as input. When used for compression, the trained CNN is applied at the encoder side, and the branch that produces the best prediction signal is encoded as side information for decoder. They integrate the method into JPEG, without changing the coding of luma, and experimental results show that the proposed method outperforms JPEG for chroma coding.

Li \emph{et al.} \cite{Li2018} propose a cross-channel prediction method analogous to LM. In particular, they design a hybrid neural network consisting of a fully connected part and a convolutional part. The former is used to process the context, including three channels, and the latter is to process the luma channel of the current block. Twofold features are fused to get the final prediction. This method outperforms LM by providing more than 2\% BD-rate for chroma coding.
\subsection{Probability Distribution Prediction}
As mentioned before, accurate probability estimation is the key problem in entropy coding. Thus, several works have been done to utilize deep learning for probability distribution prediction to improve the entropy coding efficiency. These works deal with different parts of the information. For example, the intra prediction mode of each block is required to be sent to decoder, and Song \emph{et al.} \cite{song2017neural} design a CNN to predict the probability distribution of the intra prediction mode based on the context. Similarly, Pfaff \emph{et al.} \cite{pfaff2018neural} predict the probability distribution of the intra prediction mode based on the context, but using a fully connected network. If an encoding/decoding scheme allows multiple transforms and each block can be assigned a transform mode, then Puri \emph{et al.} \cite{puri2017cnn} propose to use a CNN to predict the probability distribution of the transform mode, which is based on the quantized transform coefficients. In a more recent work, Ma \emph{et al.} \cite{Ma2018} consider the entropy coding of the quantized transform coefficients, specifically the DC coefficients. They design a CNN to predict the probability distribution of the DC coefficient of a block, from the context of the block as well as the AC coefficients of the block.
\subsection{Transform}
Transform is an important tool in the hybrid video coding framework to convert signal (usually residues) into coefficients that are then quantized and coded. At the very beginning, video coding schemes adopt discrete cosine transform (DCT), which is then replaced by integer cosine transform (ICT) in H.264. HEVC also adopts ICT but additionally uses integer sine transform for 4$\times$4 luma blocks. Adaptive multiple transforms and secondary transform are also studied. Nonetheless, all these transforms are still very simple.

Inspired by auto-encoder, Liu \emph{et al.} \cite{liu2018cnn} propose a CNN-based method to achieve a DCT-like transform for image coding. The proposed transform consists of a CNN and a fully connected layer, where the CNN is to preprocess the input block and the fully connected layer is to fulfill the \emph{transform}. In their implementation, the fully connected layer is initialized by the transform matrix of DCT, but then is trained together with the CNN. They use a joint rate-distortion cost to train the network, where rate is estimated by the $l_1$-norm of the quantized coefficients. They also investigate asymmetric auto-encoders, i.e. the encoding part and decoding part are not symmetric, different from the traditional auto-encoders. Their experimental results show that the trained transform is better than the fixed DCT, and the asymmetric auto-encoders can be useful to achieve a tradeoff between compression efficiency and encoding/decoding time.
\subsection{Post- or In-Loop Filtering}\label{sec_deeptool_filter}
Most of the widely used image and video coding schemes are lossy coding ones, i.e. the reconstructed image/video is not exactly the original image/video, for the sake of compression. The loss is usually due to the quantization process shown in Fig. \ref{fig:deep_tools}. When the quantization step is large, the loss is large too, which may lead to visible artifacts in the reconstructed image/video, such as blocking, blurring, ringing, color shift, and flickering. Filtering is the tool to reduce these artifacts, to improve the quality of the reconstructed image/video, and thus to improve the compression efficiency indirectly. For image, the filtering is also known as post-processing because it does not change the encoding process. For video, the filtering is divided into in-loop and out-of-loop, depending on whether the filtered frame is used as reference for the following frames. In HEVC, two in-loop filters are presented, namely deblocking filter (DF) \cite{norkin2012hevc} and sample adaptive offset (SAO) \cite{fu2012sample}.

Post- or in-loop filtering occupies the majority of the related works about deep learning-based image/video coding:
\begin{itemize}
\item Earlier works have focused on post-filtering for image coding, especially JPEG. For example, Dong \emph{et al.} \cite{dong2015compression} propose a 4-layer CNN for compression artifacts reduction, namely ARCNN. ARCNN achieves more than 1dB improvement in PSNR than JPEG on the 5 classical test images when the quality factor (QF) is between 10 and 40. Cavigelli \emph{et al.} \cite{cavigelli2017cas} use a deeper CNN (12-layer) with hierarchical skip connections and test for higher QF from 40 to 76. Wang \emph{et al.} \cite{Wang_2016_CVPR} leverage the prior knowledge of JPEG compression, i.e. quantization of the DCT coefficients of 8$\times$8 blocks, and propose a dual-domain (pixel domain and transform domain) based method. They achieve both higher quality and less computing time than ARCNN. Dual-domain processing is also studied in \cite{guo_building_2016}. Guo and Chao \cite{Guo_2017_CVPR} propose a one-to-many network, which is trained by a combination of perceptual loss, naturalness loss, and JPEG loss. Another work about loss function is presented in \cite{Galteri_2017_ICCV}, which suggests the usage of discriminator loss like in GAN.
Ororbia \emph{et al.} \cite{ororbia2018learned} propose an iterative post-filtering method by using a trained RNN.
Recently, several works treat JPEG post-filtering as an image restoration task, like denoising or super-resolution, and propose different networks for a series of image restoration tasks \cite{Liu_2018_CVPR_Workshops,Yu_2018_CVPR,zhang2017beyond,zhang2018adaptive}.
\item Later on, researches are more and more conducted for out-of-loop filtering in video coding, especially HEVC.
Dai \emph{et al.} \cite{dai2017convolutional} propose a 4-layer CNN for post-filtering of intra frames, where the CNN has variable filter size and residue connection, and named VRCNN.
Wang \emph{et al.} \cite{wang2017novel} use a 10-layer CNN for out-of-loop filtering, where they train a CNN to filter one image and used the trained CNN on the video frames individually.
Yang \emph{et al.} \cite{yang2017enhancing} propose to train different CNN models for I frames and P frames respectively, and verify the benefit.
Jin \emph{et al.} \cite{jin2018quality} suggest the use of a discriminator loss in addition to the MSE loss.
Li \emph{et al.} \cite{li2017cnn} propose to transmit some side information to decoder to select one model for each frame from a previously trained set of models.
In addition, Yang \emph{et al.} \cite{yang2018multi} propose to utilize the inter-picture correlation during the post-filtering process by inputting multiple neighboring frames into the CNN to enhance one frame. Wang \emph{et al.} \cite{wang2018multi} also consider the inter-picture correlation, but using a multi-scale convolutional LSTM.
While the aforementioned works take only the decoded frames as input to the CNN, He \emph{et al.} \cite{he2018enhancing} propose to input the block partition information together with decoded frame into the CNN,
Kang \emph{et al.} \cite{kangmulti} also input the block partition information into the CNN and design a multi-scale network,
Ma \emph{et al.} \cite{ma2018residual} input the intra prediction signal and the decoded residual signal into the CNN, and Song \emph{et al.} \cite{song2018practical} input the QP plus the decoded frame into the CNN (they also quantize the network parameters to ensure consistency between different computing platforms).
A different work is presented in \cite{tsai2018learning}, which does not enhance the decoded frames directly; instead, they propose to calculate the compression residues (i.e. the original video minus the decoded video, to be distinguished from prediction residues) at the encoder side, and train an auto-encoder to encode the compression residues and send to the decoder side. Their method is reported to perform well on domain-specific video sequences, e.g. in video game streaming services.
\item It is more challenging to integrate CNN-based filter into the coding loop because the filtered frame will serve as reference and will affect the other coding tools.
Park and Kim \cite{park2016cnn} train a 3-layer CNN as an in-loop filter for HEVC. They train two models for two QP ranges: 20--29 and 30--39, respectively, and use one model for each frame according to its QP. The CNN is applied after DF, and SAO is turned off. They also design two cases to apply the CNN-based filter: in the one case, the filter is applied on specified frames based on picture order count (POC); in the other, the filter is tested for each frame and if it improves quality then it is applied, one binary flag for each frame is signaled to decoder in this case.
Meng \emph{et al.} \cite{meng2018new} use an LSTM as an in-loop filter, which is applied after DF and before SAO in HEVC. The network has decoded frame together with block partition information as its input, and is trained with a combination of MS-SSIM loss and MAE loss.
Zhang \emph{et al.} \cite{zhang2018residual} propose a residual highway CNN (RHCNN) for in-loop filtering in HEVC. RHCNN-based filter is applied after SAO. They train different RHCNN models for I, P, and B frames, respectively. They also divide QPs into several ranges and train a separate model for each range.
Dai \emph{et al.} \cite{dai2018cnn} propose a deep CNN called VRCNN-ext for in-loop filtering in HEVC. They design different strategies for I frames and P/B frames: CNN-based filter replaces DF and SAO for I frames, but is applied after DF and before SAO for P/B frames with CTU- and CU-level control. At CTU-level, one binary flag for each CTU is signaled to control the on/off of CNN-based filter; if the flag is off, then at CU-level, a binary classifier is used to decide whether to turn on CNN-based filter for each CU.
Jia \emph{et al.} \cite{jia2019content} also consider a deep CNN for in-loop filtering in HEVC. The filter is applied after SAO and controlled by frame- and CTU-level flags. If frame-level flag is ``off,'' then the corresponding CTU-level flags are omitted. In addition, they train multiple CNN models and train a content analysis network that decides one model for each CTU, which saves the bits of CNN model selection.
\end{itemize}
\subsection{Down- and Up-Sampling}
A trend of the video technology is to increase the resolution at different dimensions, such as spatial resolution (i.e. number of pixels), temporal resolution (i.e. frame rate), and pixel value resolution (i.e. bit-depth).
The increasing resolution results in multiplied data volume, which raises a great challenge to video transmission systems.
When the bandwidth for transmission is limited (e.g. using 2G or 3G mobile network), a common practice is to decrease video resolution before encoding and to increase video resolution back after decoding.
This is known as the down- and up-sampling-based coding strategy.
The down- and up-sampling can be performed in the spatial domain, the temporal domain, the pixel value domain, or a combination of these domains.
Traditionally, the down- and up-sampling filters are often handcrafted.
Recently, it is proposed to train deep networks as down- and up-sampling filters for efficient video coding.
There are two categories of related researches.

The first category is focused on training deep networks as up-sampling filters only, while still using handcrafted down-sampling filters. This is inspired by the success of super-resolution, e.g. \cite{dong2014learning}.
For example in \cite{afonso2019video}, a joint spatial and pixel value down-sampling is proposed, where the spatial down-sampling is achieved by a handcrafted low-pass filter and the pixel value down-sampling is achieved by bitwise right shift. At the encoder side, a support vector machine is used to decide whether to perform down-sampling for each frame. At the decoder side, a CNN is trained to up-sample the decoded video to its original resolution.
In \cite{li2018convolutional}, Li \emph{et al.} only consider spatial down-sampling which is also performed by a handcrafted filter, and train a CNN for up-sampling. But different from \cite{afonso2019video}, they propose a block adaptive resolution coding (BARC) framework. Specifically, for each block inside a frame, they consider two coding modes: down-sampling then coding and direct coding. The encoder can choose a mode for each block and signal the chosen mode to the decoder. In addition, in the down-sampling coding mode, they further design two sub-modes: using a handcrafted simple filter for up-sampling, and using the trained CNN for up-sampling. The sub-mode is also signaled to the decoder.
Li \emph{et al.} \cite{li2018convolutional} investigate BARC only for I frames.
Later, Lin \emph{et al.} \cite{lin2018convolutional} extend the BARC framework for P and B frames and build a complete BARC-based video coding scheme.
While the aforementioned works perform down-sampling in the pixel domain, Liu \emph{et al.} \cite{liu2018convolutional} propose down-sampling in the residue domain, i.e. they down-sample the inter prediction residues, and they up-sample the residues by a trained CNN with considering the prediction signal. They also follow the BARC framework.

The second category trains not only up-sampling but also down-sampling filters to allow for more flexibility.
For example in \cite{jiang2017end}, a compression framework with two CNNs is studied. The first CNN down-samples an image, the down-sampled image is then compressed by an existing image encoder (such as JPEG and BPG), and then decoded, the second CNN up-samples the decoded image.
One drawback of this framework is that it cannot be trained end-to-end because the image encoder/decoder is not differentiable.
To address this problem, Jiang \emph{et al.} \cite{jiang2017end} decide to optimize the two CNNs alternatively.
Differently, Zhao \emph{et al.} \cite{zhao2017learning} use a virtual codec that is actually a CNN to approximate the functionality of--and thus replace--the encoder/decoder; they also insert a CNN to perform post-processing before the up-sampling CNN; their scheme is fully convolutional and can be trained end-to-end.
Moreover, Li \emph{et al.} \cite{li2019learning} simply remove the encoder/decoder and keep only the two CNNs during training; considering that the down-sampled image will be compressed, they propose a novel regularization loss for training, which requires the down-sampled image to be not quite different from the ideal low-pass and decimated (which is approximated by a handcrafted filter) image. The regularization loss is verified to be useful when training down- and up-sampling CNNs jointly for image coding.
\subsection{Encoding Optimizations}
The aforementioned deep tools are intended for increasing the compression efficiency, especially for reducing bitrate while keeping the same PSNR.
There are some other deep tools that target different aspects.
In this subsection, we review several deep tools for three different objectives: fast encoding, rate control, and region-of-interest (ROI) coding. Since these tools are used only at the encoder side, we call them encoding optimization tools in summary.
\subsubsection{Fast Encoding}
Regarding the state-of-the-art video coding standards, H.264 and HEVC, the decoder is computationally simple, but the encoder is much more complex. This is because more and more coding modes are introduced into the video coding standards, and each block can be assigned a different mode. The mode of each block is signaled to the decoder, so the decoder only needs to compute the given mode. But to find the mode for each block, the encoder usually needs to compare the multiple optional modes and select the optimal one, where optimality is claimed in the rate-distortion sense. Therefore, if the encoder performs an exhaustive search, then the compression efficiency is the highest, but the computational complexity may be also very high. Any practical encoder will adopt heuristic algorithms to search for a better mode, where machine learning especially deep learning can help.

Liu \emph{et al.} \cite{liu2016cu} present a hardware design for HEVC intra encoder, where they adopt a trained CNN to help decide CU partition mode. Specifically in HEVC intra coding, a CTU is split into CUs recursively to form a quadtree structure. Their trained CNN will decide whether to split a 32$\times$32/16$\times$16/8$\times$8 CU or not based on the content inside the CU and the specified QP. Actually, this is a binary decision problem.
Xu \emph{et al.} \cite{xu2017reducing} additionally consider HEVC inter encoder, and propose an early-terminated hierarchical CNN and an early-terminated hierarchical LSTM to help decide CU partition mode, for I frames and P/B frames, respectively.
Jin \emph{et al.} \cite{jin2017cnn} also consider the CU partition mode decision but for the incoming VVC rather than HEVC, because in VVC a quadtree-bintree (QTBT) structure is designed for CU partition, which is more complex than that in HEVC. They train a CNN to perform 5-way classification for a 32$\times$32 CU, where different classes indicate different tree depths.
Xu \emph{et al.} \cite{xu2018fast} investigate the CU partition mode decision for H.264 to HEVC transcoding. They design a hierarchical LSTM network to predict the CU partition mode from the features extracted from H.264 coded bits.

Song \emph{et al.} \cite{song2017cnn} study a CNN-based method for fast intra prediction mode decision in HEVC intra encoder. They train a CNN to derive a list of most probable modes for each 8$\times$8/4$\times$4 block based on the content and the specified QP, and then choose a mode from the list by the normal rate-distortion optimized process.
\subsubsection{Rate Control}
Given a limited transmission bandwidth, video encoder tries to produce bits that do not overflow the bandwidth. This is known as the rate control requirement.

One traditional rate control method is to allocate bits to different blocks according to the R-$\lambda$ model \cite{li2014lambda}. In that model, each block has two parameters $\alpha$ and $\beta$ that are to be determined. Previously, the parameters are estimated by an empirical formula. In \cite{li2017convolutional}, Li \emph{et al.} propose to train a CNN to predict the parameters for each CTU. Experimental results show that the proposed method achieves higher compression efficiency as well as lower rate control error.

Hu \emph{et al.} \cite{hu2018reinforcement} attempt to leverage reinforcement learning methods for intra-frame rate control. They draw an analogy between the rate control problem and the reinforcement learning problem: the texture complexity of the blocks and bit balance are regarded as the environment state, the quantization parameter is regarded as an action that an agent needs to take, and the negative distortion of the blocks is regarded as an immediate reward. They train a neural network as the agent.
\subsubsection{ROI Coding}
ROI refers to the regions of interest in an image. In image compression, it is often required that the content in ROI shall be of high quality and the content not in ROI can be of low quality. Many image coding schemes, such as JPEG and JPEG 2000, support ROI coding. Then, how to identify the ROI is a research problem and has been addressed by using deep learning.
Prakash \emph{et al.} \cite{prakash2017semantic} propose a CNN-based method to generate a multi-scale ROI (MS-ROI) map to guide the following JPEG coding. They use a trained image classification network on an image, pick the top five classes predicted by the network, and identify the regions that correspond to these classes. Thus, their MS-ROI map indicates salient regions that are related to semantic analysis.
\section{Case Study of DLVC}\label{sec_dlvc}
\begin{figure*}[tb]
\centering
\includegraphics[width=1.0\linewidth]{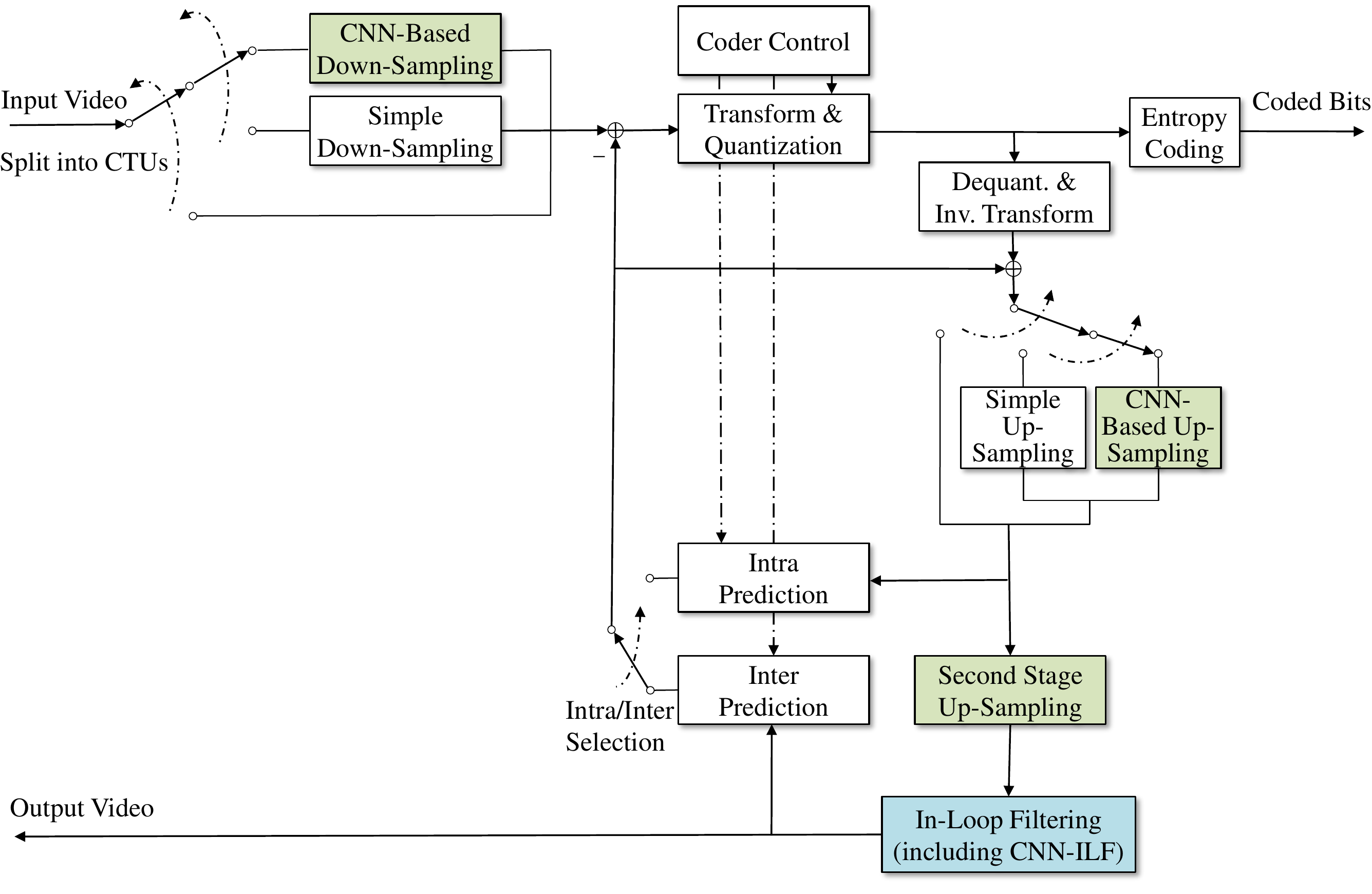}
\caption{Illustration of the developed DLVC scheme. Inside the blue block is the proposed CNN-ILF. The green blocks correspond to CNN-BARC.}
\label{fig_framework}
\end{figure*}
We now turn to the case study of our developed DLVC, a prototype video codec. Indeed, DLVC was developed as a proposal in response to the joint call for proposals on video compression with capability beyond HEVC. Now the source code of DLVC has been released for future researches\footnote{\url{https://github.com/fvc2018/dlvc}, \url{http://dlvc.bitahub.com/}.}. DLVC is crafted upon the JEM software, contains a number of improvements than JEM, and especially has two deep coding tools: CNN-based in-loop filter (CNN-ILF) and CNN-based block adaptive resolution coding (CNN-BARC), both of which are based on trained CNN models. The scheme of DLVC is illustrated in Fig. \ref{fig_framework}. In this section, we focus on the two deep tools. More technical details about DLVC can be found in the technical report \cite{Wu2018Description}.
\subsection{CNN-Based In-Loop Filter}
\label{sec_CNNLF}
As mentioned in Section \ref{sec_deeptool_filter}, a great number of researches have been conducted on using trained CNN models for post- or in-loop filtering. CNN-ILF represents our efforts at this aspect.

\begin{figure*}[tb]
\centering
\includegraphics[width=0.6\linewidth]{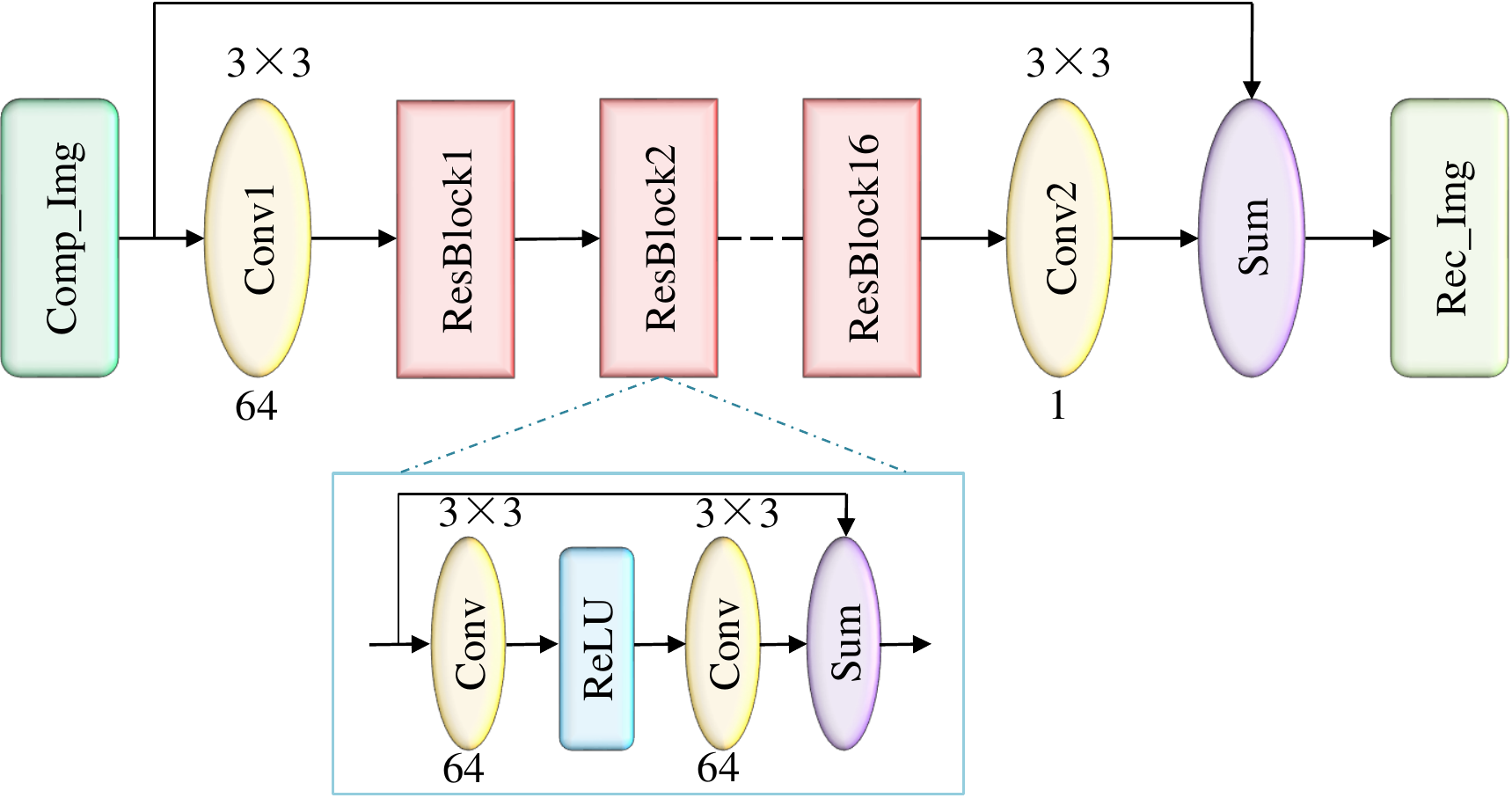}
\caption{The network structure of the proposed CNN-ILF. The numbers shown above and beneath each convolutional (Conv) layer indicate the kernel size (e.g. 3$\times$3) and the number of output channels (e.g. 64), respectively.}
\label{fig_CNN-LF}
\end{figure*}
The network structure of our proposed CNN-ILF is illustrated in Fig. \ref{fig_CNN-LF}.
Inspired by the SR network in \cite{lim2017enhanced}, we design a deep CNN having 16 residual blocks (ResBlocks) and 2 convolutional layers, in total 34 layers. Each ResBlock consists of 2 convolutional layers separated by a ReLU mapping, and a skip connection. The entire network has a global skip connection from the input to the output. These skip connections are crucial to train an efficient network and to accelerate the convergence in training.

To train the network, we have used a set of natural images and compressed each image by the DLVC intra coding (turning off all in-loop filters) at different QPs. For each QP we train a separate model. We only use the luma component for training but the trained models are used for both luma and chroma channels during compression. We divide images into 70$\times$70 sub-images and shuffle the sub-images to prepare training data. The loss function is MSE, i.e. the error between the network-output image and the original uncompressed image. We use the stochastic gradient descent algorithm to train the network until convergence.

We apply the trained models in DLVC. The CNN-ILF is applied after deblocking filter and before sample adaptive offset.
There are different models corresponding to different QPs, and one model is selected for each frame according to the frame's QP.
For each CTU there are two binary flags that control the on/off of the CNN-ILF for luma and chroma respectively. These flags are decided at the encoder side and transmitted to the decoder.
\subsection{CNN-Based Block Adaptive Resolution Coding}
\label{sec_CNN-BARC}
CNN-BARC is a down- and up-sampling-based coding tool that uses trained CNN models as the down- and up-sampling filters. In DLVC, CNN-BARC is applied only for intra frame coding.
The down-sampling coding or direct coding mode is decided for each CTU, and the down-sampling coding mode has two sub-modes: using CNNs for down- and up-sampling, and using simple interpolation filters for down- and up-sampling. All the modes and sub-modes are decided by the encoder and signaled to the decoder.

\begin{figure*}[tb]
  \centering
  \subfigure[]{\includegraphics[width=0.8\linewidth]{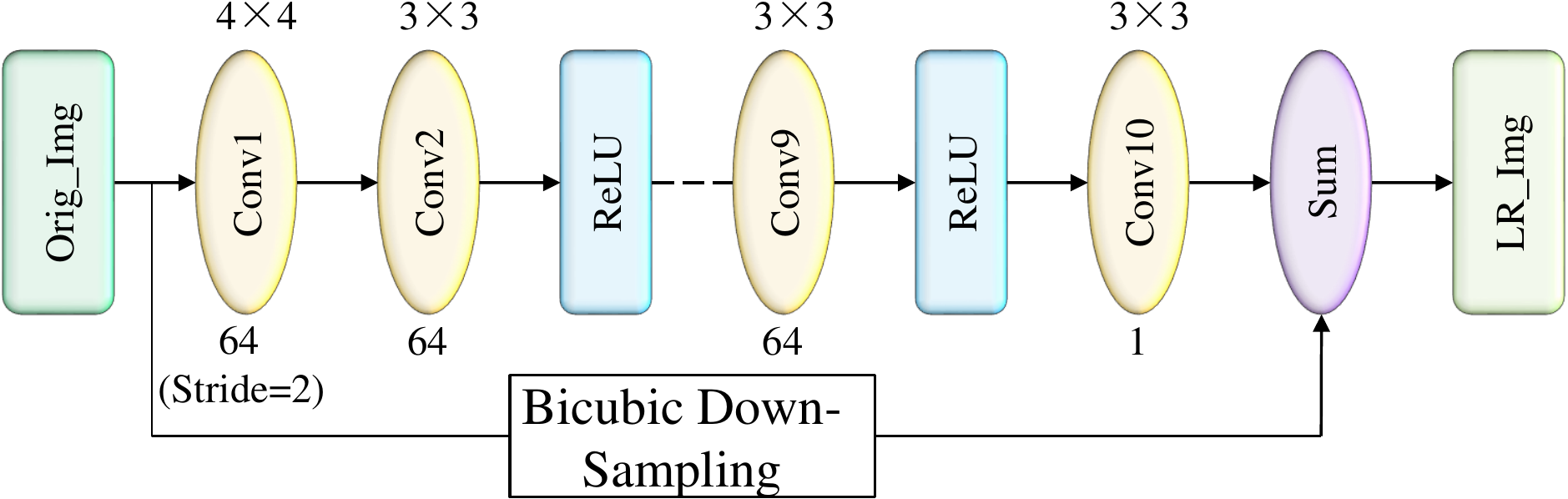}}
  \subfigure[]{\includegraphics[width=0.8\linewidth]{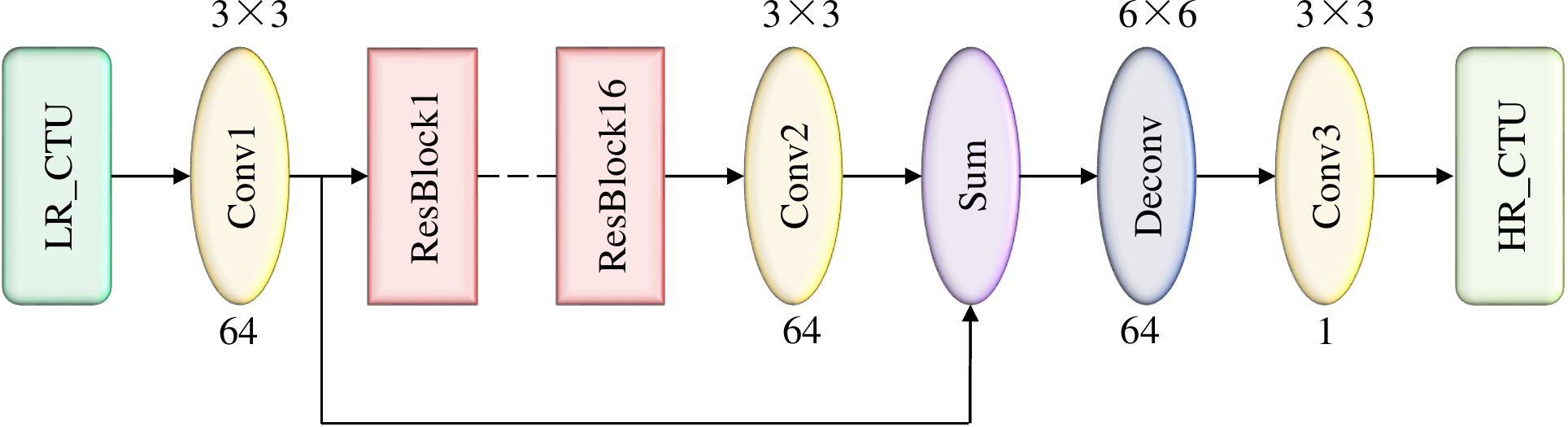}}
  \caption{The network structure of the proposed CNN-BARC, including (a) CNN-DS for down-sampling and (b) CNN-US for up-sampling. Note that the first Conv layer in (a) has a stride of 2 to achieve 2$\times$ down-sampling.}\label{fig_Down_Up_Sampling_CNN}
\end{figure*}
The networks for down- and up-sampling are illustrated in Fig. \ref{fig_Down_Up_Sampling_CNN}. Specifically, the down-sampling CNN (CNN-DS) has 10 convolutional layers where the first layer has a stride of 2 to achieve 2$\times$ down-sizing. CNN-DS also embraces residue learning but here the original image is bicubic down-sampled to serve as the skip connection.
The up-sampling CNN (CNN-US) is similar to the SR network in \cite{lim2017enhanced}, and has 16 ResBlocks, 3 convolutional layers, 1 deconvolutional layer, and a global skip connection.

The CNN-DS and CNN-US are trained in four steps. First, we remove the convolutional layers in the CNN-DS, making it a simple bicubic down-sampling, and train the CNN-US to minimize the end-to-end MSE (i.e. the error between original image and down-sampled-up-sampled image). Second, we add back the layers of CNN-DS, fix the parameters of the CNN-US, and train the CNN-DS to minimize the end-to-end MSE. Third, we fine-tune the parameters of CNN-DS and CNN-US simultaneously, using a combination of two losses: the one is end-to-end MSE, and the other is the down-sampled MSE (i.e. the error between bicubic down-sampled image and network down-sampled image), where the latter serves as a regularization term. Fourth, we fix the parameters of the CNN-DS, and compress the down-sampled images by the DLVC intra coding (turning off all in-loop filters) at different QPs. For each QP we train a separate CNN-US model.

There are two mode selection steps regarding CNN-BARC in the DLVC encoder. The first is to decide which down- and up-sampling (sub-)mode, and the second is to decide whether to perform down-sampling. We compare the rate-distortion costs of different modes to make decision. The rate is the number of coded bits, and the distortion is the MSE between original and reconstructed CTUs. For fair comparison, we always calculate the distortion at the original resolution. Last but not the least, after an intra frame is compressed, we perform up-sampling again for the down-sampled-coded CTUs. More details about CNN-BARC can be found in \cite{li2019learning,li2018convolutional}.
\subsection{Compression Performance}
\begin{table}[tb]
\centering
\caption{BD-rate results of DLVC compared to HM16.16}\label{Tab:DLVC_anchor}
\begin{tabular}{l|c|c|c|c|c|c}
\hline
 &
\multicolumn{3}{c|}{Random-Access} &
\multicolumn{3}{c}{Low-Delay}\\
\cline{2-7}
   & Y & U & V & Y & U & V\\
 \hline\hline
 FoodMarket &$-$38.42\% &$-$45.57\% &$-$50.10\% &- &- &-\\
 CatRobot &$-$46.87\% &$-$59.26\% &$-$53.80\% &- &- &-\\
 DaylightRoad &$-$47.34\% &$-$60.65\% &$-$46.19\% &- &- &-\\
 ParkRunning &$-$38.43\% &$-$29.65\% &$-$31.58\% &- &- &-\\
 CampfireParty &$-$41.20\% &$-$38.07\% &$-$54.53\% &- &- &-\\
 \hline
 Avg. Class-A &$-$42.45\% &$-$46.64\% &$-$47.24\%  &- &- &- \\
\hline\hline
 BQTerrace &$-$37.53\% &$-$50.06\% &$-$56.97\% &$-$35.56\% &$-$52.79\% &$-$59.11\%\\
 RitualDance &$-$33.30\% &$-$42.15\% &$-$45.71\% &$-$28.23\% &$-$41.19\% &$-$43.19\%\\
 MarketPlace &$-$35.34\% &$-$51.78\% &$-$49.99\% &$-$26.85\% &$-$46.65\% &$-$47.83\%\\
 BasketballDrive &$-$37.09\% &$-$52.86\% &$-$51.20\% &$-$32.98\% &$-$50.91\% &$-$52.69\%\\
 Cactus &$-$40.80\% &$-$50.92\% &$-$46.82\% &$-$41.37\% &$-$56.66\% &$-$54.68\%\\
\hline
Avg. Class-B &$-$36.81\% &$-$49.55\% &$-$50.14\%  &$-$33.00\% &$-$49.64\% &$-$51.50\% \\
\hline\hline
\textbf{Avg. All}  &\textbf{$-$39.63\%} &\textbf{$-$48.10\%} &\textbf{$-$48.69\%}  &\textbf{$-$33.00\%} &\textbf{$-$49.64\%} &\textbf{$-$51.50\%} \\
\hline
\end{tabular}
\end{table}

\begin{table}[tb]
\centering
\caption{BD-rate results of all the proposals in response to the joint call for proposals compared to JEM7.0}
\begin{tabular}{l|l|r|r|r|r|r|r}
 \hline
 \multirow{2}{*}{Proposal} &\multirow{2}{*}{Organizations}&\multicolumn{3}{c|}{Random-Access}&\multicolumn{3}{c}{Low-Delay}\\
 \cline{3-8}
 & & \multicolumn{1}{c|}{Y} & \multicolumn{1}{c|}{U} & \multicolumn{1}{c|}{V} & \multicolumn{1}{c|}{Y} & \multicolumn{1}{c|}{U} & \multicolumn{1}{c}{V}\\
 \hline\hline
  J0011&DJI, Peking Univ.          &$-$1.57\%&$-$0.71\%&$-$1.72\%  &$-$3.30\% &$-$0.67\% &$-$4.26\% \\
\hline
 J0012&Ericsson, Nokia            &$-$0.90\%&	$-$1.14\%&$-$1.15\% &$-$0.17\% &$-$0.48\%&$-$1.93\% \\
\hline
 J0013&ETRI, Sejong Univ.            &0.64\%&	$-$0.39\%&$-$0.89\% &0.85\% &$-$1.27\%&$-$2.50\% \\
\hline
J0014&Fraunhofer HHI            &$-$7.55\%&	$-$6.94\%&$-$5.96\% &$-$7.22\% &$-$7.62\%&$-$5.71\% \\
\hline
\multirow{2}{*}{J0015}&\multirow{2}{*}{InterDigital, Dolby}   &\multirow{2}{*}{$-$3.98\%}&	\multirow{2}{*}{$-$3.28\%}&\multirow{2}{*}{$-$3.16\%} &\multirow{2}{*}{$-$3.64\%} &\multirow{2}{*}{$-$2.55\%}&\multirow{2}{*}{$-$4.48\%} \\
 & & & & & & & \\
 \hline
 J0016&KDDI            &$-$0.57\%&	$-$0.52\%&$-$1.30\% &$-$0.17\% &0.38\%&$-$0.40\% \\
 \hline
 J0017&LG Electronics            &$-$2.52\%&	$-$5.29\%&$-$6.19\% &$-$2.09\% &$-$2.47\%&$-$3.70\% \\
 \hline
 \multirow{2}{*}{J0018}&\multirow{2}{*}{MediaTek}   &$-$16.60\%&	$-$6.75\%&$-$10.43\% &$-$9.41\% &$-$1.92\%&$-$3.35\% \\
 \cline{3-8}
 & &$-$14.40\%&	$-$5.13\%&$-$8.82\% &$-$7.20\% &0.23\%&$-$0.96\% \\
 \hline
 \multirow{2}{*}{J0020}&\multirow{2}{*}{Panasonic}   &$-$2.28\%&	$-$3.44\%&$-$3.88\% &$-$2.02\% &$-$1.02\%&$-$2.36\% \\
 \cline{3-8}
 & &$-$0.06\%&	0.91\%&0.52\% &$-$0.09\% &2.87\%&1.21\% \\
 \hline
 \multirow{2}{*}{J0021}&\multirow{3}{*}{Qualcomm, Technicolor}   &$-$15.53\%&	$-$3.66\%&$-$5.97\% &$-$12.65\% &$-$17.40\%&$-$19.95\% \\
 \cline{3-8}
 & &$-$10.26\%&	0.05\%&$-$1.65\% &$-$9.87\% &$-$11.73\%&$-$14.88\% \\
  \cline{1-1}\cline{3-8}
 J0022& &$-$13.60\%&$-$3.80\%&$-$5.63\% &$-$12.69\% &$-$16.65\%&$-$18.75\% \\
 \hline
 J0023&RWTH Aachen Univ.            &$-$0.79\%&	$-$1.52\%&$-$1.52\% &$-$0.84\% &$-$0.58\%&$-$0.80\% \\
\hline
\multirow{2}{*}{J0024}&Samsung, Huawei,  &$-$6.01\%&	10.34\%&8.53\% &$-$0.38\% &14.73\%&15.84\% \\
 \cline{3-8}
 &GoPro, HiSilicon &$-$4.24\%&	10.71\%&9.23\% &$-$ &$-$&$-$ \\
 \hline
 J0026&Sharp, Foxconn            &$-$6.15\%&	$-$5.68\%&$-$5.68\% &$-$5.57\% &9.00\%&8.69\% \\
\hline
J0027&NHK, Sharp            &$-$2.14\%&	$-$5.55\%&$-$5.61\% &$-$2.23\% &$-$1.97\%&$-$3.83\% \\
\hline
J0028&Sony            &$-$2.41\%&	$-$4.85\%&$-$5.14\% &$-$2.25\% &$-$6.74\%&$-$7.34\% \\
\hline
J0029&Tencent            &$-$4.70\%&	$-$8.34\%&$-$8.91\% &$-$4.47\% &$-$9.47\%&$-$10.82\% \\
\hline
J0031&Univ. Bristol            &$-$4.54\%&	20.19\%&18.68\% &$-$0.52\% &5.74\%&5.81\% \\
\hline
\textbf{J0032}&\textbf{USTC and others}     &\textbf{$-$10.11\%}&\textbf{	$-$9.59\%}&\textbf{$-$9.97\%} &\textbf{$-$11.83\%} &\textbf{$-$13.22\%}&\textbf{$-$16.49\%} \\
\hline
\end{tabular}
\\
\footnotesize\selectfont
More details can be obtained at \url{http://phenix.it-sudparis.eu/jvet/doc_end_user/current_meeting.php?id_meeting=174&search_id_group=1&search_sub_group=1}.
\label{Tab:Cfp_submissions}
\end{table}

\begin{table}[tb]
\centering
\caption{BD-rate results of CNN-ILF on top of HM16.16 + QTBT} \label{Tab:CNNLF_anchor}
\begin{tabular}{c|c|c|c|c|c|c|c|c|c}
\hline
 &
\multicolumn{3}{c|}{Random-Access} &
\multicolumn{3}{c|}{Low-Delay}&
\multicolumn{3}{c}{All-Intra}\\
\cline{2-10}
   & Y & U & V & Y & U & V& Y & U & V\\
\hline
Avg. Class-A  &$-$5.1\% &$-$3.8\% &$-$4.5\%  &- &- &- &$-$5.7\% &$-$5.1\% &$-$5.9\% \\
\hline
Avg. Class-B  &$-$5.9\% &$-$5.8\% &$-$5.8\%  &$-$5.2\% &$-$7.2\% &$-$8.4\% &$-$7.1\% &$-$6.5\% &$-$7.8\% \\
\hline
\textbf{Avg. All}  &\textbf{$-$5.5\%} &\textbf{$-$4.8\%} &\textbf{$-$5.1\%}  &\textbf{$-$5.2\%} &\textbf{$-$7.2\%} &\textbf{$-$8.4\%} &\textbf{$-$6.4\%} &\textbf{$-$5.8\%} &\textbf{$-$6.8\%}\\
\hline
\end{tabular}
\end{table}

\begin{table}[tb]
\centering
\caption{BD-rate results of CNN-BARC on top of HM16.16 + QTBTTT} \label{Tab:CNN-BARC_anchor}
\begin{tabular}{c|c|c|c|c|c|c|c|c|c}
\hline
 &
\multicolumn{3}{c|}{Random-Access} &
\multicolumn{3}{c|}{Low-Delay}&
\multicolumn{3}{c}{All-Intra}\\
\cline{2-10}
   & Y & U & V & Y & U & V& Y & U & V\\
\hline
Avg. Class-A  &$-$1.8\% &0.6\% &0.4\%  &- &- &- &$-$7.2\% &2.0\% &2.0\% \\
\hline
Avg. Class-B  &$-$1.0\% &0.1\% &$-$0.1\%  &$-$0.3\% &$-$0.3\% &$-$0.0\% &$-$3.6\% &0.5\% &0.4\% \\
\hline
\textbf{Avg. All}  &\textbf{$-$1.4\%} &\textbf{0.3\%} &\textbf{0.2\%}  &\textbf{$-$0.3\%} &\textbf{$-$0.3\%} &\textbf{$-$0.0\%} &\textbf{$-$5.4\%} &\textbf{1.3\%} &\textbf{1.2\%}\\
\hline
\end{tabular}
\end{table}
We test the compression performance of DLVC on the 10 video sequences that are recommended by JVET. These sequences are divided into Class A and Class B according to spatial resolution: 5 sequences have UHD (3840$\times$2160) resolution and the other 5 have HD (1920$\times$1080) resolution. Different coding configurations including all-intra (AI), low-delay (LD), and random-access (RA) are tested. We compare DLVC with the HEVC reference software--HM version 16.16\footnote{HM version 16.16, \url{https://hevc.hhi.fraunhofer.de/svn/svn_HEVCSoftware/tags/HM-16.16/}.} and its variants as well as JEM version 7.0\footnote{JEM version 7.0, \url{https://jvet.hhi.fraunhofer.de/svn/svn_HMJEMSoftware/tags/HM-16.6-JEM-7.0/}.}, and use BD-rate \cite{bjontegaard2001calcuation} to measure the relative compression efficiency.

Table \ref{Tab:DLVC_anchor} presents the BD-rate results of DLVC compared to the HEVC anchor. Obviously, DLVC improves the compression efficiency very significantly.
Considering the Y channel, DLVC achieves on average 39.6\% and 33.0\% BD-rate reduction than HEVC, under RA and LD configurations, respectively. The results indicate that when using DLVC to replace HEVC, the bits can be reduced by more than 30\% without lowering the reconstruction quality.

Table \ref{Tab:Cfp_submissions} presents the BD-rate results of DLVC (document number J0032) compared to the JEM anchor. For comparison, we also include the BD-rate results of the other proposals in response to the joint call for proposals. Considering the Y channel, DLVC achieves on average 10.1\% and 11.8\% BD-rate reduction than JEM, under RA and LD configurations, respectively. DLVC is among the best proposals from the BD-rate perspective.

Table \ref{Tab:CNNLF_anchor} verifies the effectiveness of the proposed CNN-ILF. Specifically, we use a variant of HM that adds the QTBT structure, which outperforms the vanilla HM, as the anchor. We integrate the CNN-ILF into the anchor and turn on/off the CNN-ILF for comparison. As shown, CNN-ILF achieves significant BD-rate reduction: on average 5.5\%, 5.2\%, 6.4\% for the Y channel under RA, LD, AI configurations, respectively.

Table \ref{Tab:CNN-BARC_anchor} verifies the effectiveness of the proposed CNN-BARC. We use another variant of HM that adds the quadtree-bintree-triple-tree (QTBTTT) structure, which further outperforms the HM plus QTBT, as the anchor. We integrate the CNN-BARC into the anchor and turn on/off the CNN-BARC for comparison. As shown, CNN-BARC achieves significant BD-rate reduction under AI configuration: on average 5.4\% for the Y channel. The BD-rate under RA and LD configurations is less significant, because CNN-BARC is applied on intra frames only.
\section{Perspectives and Conclusions}\label{sec_concl}
\subsection{Open Problems}
\begin{itemize}
  \item \emph{Deep Schemes or Deep Tools.} Shall we be ambitious to expect deep scheme to be the future of video coding, or shall we be satisfied with deep tools within traditional non-deep schemes? In other words, can the non-deep schemes be completely replaced by deep schemes? As for now the answer to this question is probably ``no'' because deep schemes in general do not outperform non-deep schemes for video coding. But as research goes on, the answer may become ``yes'' via two ways: first, deep schemes may be improved so much that they clearly beat non-deep schemes; second, the coding tools in a traditional coding scheme (e.g. HEVC) may be all replaced by corresponding deep tools, leading to a ``deepened'' coding scheme that is better than before. The second way may be more practical according to our subjective feeling.
  \item \emph{Compression Efficiency versus Computational Complexity.} Compared the existing deep tools with their counterparts in the traditional non-deep schemes, one may easily notice that the computational complexity of the former is much higher than the latter. High complexity is indeed a general issue of deep learning, and a critical issue that hinders the adoption of deep networks in scenarios of limited computing resource, e.g. mobile phones. This general issue is now addressed at two aspects: first, to develop novel, efficient, compact deep networks that maintain the high performance (i.e. compression efficiency for video coding) but require much less computations; second, to advocate the adoption of hardware that is specifically designed for deep networks.
  \item \emph{Optimization for Perceptual Naturalness or Semantic Quality.} Coding schemes designed for natural video are usually serving for human viewing, e.g. television, movie, micro-video. It is natural for these schemes that the quality of reconstructed video shall be evaluated based on human perception. Nonetheless, for traditional non-deep coding schemes, the most widely adopted quality metric is still PSNR, which corresponds to human perception poorly. For deep schemes or deep tools, a few works have been done to optimize them for perceptual naturalness, e.g. using discriminator loss. Moreover, there are coding schemes that serve for automatic semantic analysis instead of human viewing, such as surveillance video coding. For these schemes, the quality metric shall be semantic quality \cite{liu2017recognizable}, which remains largely unexplored. As a special note, we find that there is a tradeoff between signal fidelity, perceptual naturalness, and semantic quality \cite{liu2019classification}, which implies that the optimization target shall be aligned with the actual requirement.
  \item \emph{Speciality and Universality.} To one extreme, can one coding scheme be simply the best for any kind of video? The answer is ``no'' due to the no-free-lunch theorem, which is claimed in the machine learning literature \cite{wolpert1997no} and also applies for compression. To another extreme, can we have a special coding scheme for each video? Not to mention the practical difficulty, such a coding ``strategy'' is useless because it is no more than assigning an identifier to each video. In between the two extremes are practical and useful coding schemes. That says, coding schemes shall have both speciality and universality to some extent. For deep schemes and deep tools, it implies that the training data shall be carefully selected to reflect the interesting data domain. Researches at this aspect are expected.
  \item \emph{Federated Design of Multiple Deep Tools.} Currently, most of deep tools have been designed individually, but once they are applied jointly, they may not collaborate well, or may even conflict with each other. The underlying reason is that multiple coding tools are indeed dependent. For example, different prediction tools generate different predictions and lead to variety of residual signal, so transform tools dealing with residual signal perform differently. Ideally, multiple deep tools shall be designed in a federated manner. However, this can be difficult because the dependency among tools is complicated.
\end{itemize}
\subsection{Future Work}
In the predictable future, the requirement about video coding technology is still increasing.
For entertainment, virtual reality and augmented reality applications are calling for techniques to tackle with new data, such as depth map, point cloud, 3D surface, and so on. For surveillance, the need of intelligent analysis pushes the upgrade of video resolution. For scientific observation, more and more observing instruments are directly connected to a video recorder and generate massive video data. All these requirements drive the development of video coding to achieve higher compression efficiency, lower computational complexity, and smarter integration into video analytical systems. We believe deep learning-based video coding techniques are promising for these challenging objectives. Especially, we expect a holistic framework based on deep networks and integrating image/video acquisition, coding, processing, analysis, and understanding, which indeed mimics human vision system.
\section*{Acknowledgment}
This work is supported by the Natural Science Foundation of China under Grant 61772483. Authors would like to thank the following colleagues and collaborators: Jizheng Xu, Bin Li, Zhibo Chen, Li Li, Fangdong Chen, Yuanying Dai, Changsheng Gao, Lei Guo, Shuai Huo, Ye Li, Kang Liu, Changyue Ma, Haichuan Ma, Rui Song, Yefei Wang, Ning Yan, Kun Yang, Qingyu Zhang, Zhenxin Zhang, and Haitao Yang.

\end{document}